\PassOptionsToPackage{normalem}{ulem}

\documentclass[referee]{raa}

\usepackage{graphicx,times}
\usepackage{natbib}
\usepackage{amssymb,amsmath}
\usepackage{xcolor}
\usepackage{xeCJK}
\usepackage{subcaption}
\usepackage{lscape}
\usepackage{placeins}
\usepackage{longtable}
\usepackage{graphicx}
\usepackage[colorlinks=true,linkcolor=blue,citecolor=blue,urlcolor=blue]{hyperref}

\bibpunct{(}{)}{;}{a}{}{,}

\begin{document}

   \title{The stellar orbit distribution of 24 late-type galaxies observed by MUSE}

   \volnopage{Vol.0 (20xx) No.0, 000--000}
   \setcounter{page}{1}

   \author{Le Zhang\inst{1,2}\thanks{E-mail: zhangle@shao.ac.cn}
          \and Ling Zhu\inst{1}\thanks{Corresponding author: lzhu@shao.ac.cn}
          \and
          Jes{\'u}s Falc{\'o}n-Barroso\inst{3,4}}

   \institute{Shanghai Astronomical Observatory, Chinese Academy of Sciences, 80 Nandan Road, Shanghai 200030, China\\
             \and
             School of Astronomy and Space Sciences, University of Chinese Academy of Sciences, No. 19A Yuquan Road, Beijing 100049, People's Republic of China\\
             \and
             Instituto de Astrof\'isica de Canarias, Calle Via L\'{a}ctea s/n, 38200 La Laguna, Tenerife, Spain \\
             \and
             Depto. Astrof\'isica, Universidad de La Laguna, Calle Astrof\'isico Francisco S\'{a}nchez s/n, 38206 La Laguna, Tenerife, Spain \\
             \vs\no
             {\small Received September 30, 2026}}
             
    \titlerunning{Orbital Decomposition of MUSE Galaxies}
    \authorrunning{Le Zhang et al.}
   \date{Received September 30, 2026}

    \abstract{We selected 24 nearby late-type galaxies from the MUSE archive with kinematic coverage extending to at least one effective radius. Stellar kinematics were extracted with the GIST pipeline using pPXF, and axisymmetric Schwarzschild models were constructed with \textsc{dynamite}. The stellar orbits of each galaxy were classified by circularity into dynamically cold, warm, hot, and counter-rotating components, and we calculated the corresponding luminosity fractions of each component within $R_{\mathrm{e}}$. Our results confirm the mass-dependent trends in orbital structure previously identified from CALIFA galaxies and reproduced by IllustrisTNG. We further checked the morphology of each component and calculated the edge-on flattening of the cold, warm, and hot+CR components. With the high-quality MUSE data, we find that the dynamically warm components are common in low-mass late-type spiral galaxies, suggesting that their discs are dynamically warmer and vertically thicker than those of more massive spirals. We regard this as a tentative result, given the limited sample size.
    \keywords{galaxies: kinematics and dynamics --
                galaxies: spiral --
                galaxies: structure --
                galaxies: evolution --
                galaxies: bulges
               }
    }

   \maketitle

\section{Introduction}
\label{sec:intro}

    The structural evolution of galaxies is governed by the complex interplay between hierarchical merging \citep{white1978core, white1991galaxy}, gas accretion \citep{keres2005how, dekel2009cold}, and internal secular processes. In the standard hierarchical picture, as realised in cosmological hydrodynamical simulations \citep{springel2005simulations, vogelsberger2014properties}, major mergers and accretion are important channels for building massive ellipticals and classical bulges \citep{toomre1972galactic, barnes1992dynamics}, although their relative contributions remain model-dependent \citep{hopkins2010mergers}. In contrast, late-type galaxies evolve through prolonged, steady gas accretion that builds rotationally supported discs, coupled with slow secular processes such as bar instabilities and spiral density waves that drive gas inflows to form pseudo-bulges \citep{kormendy2004secular}. To decipher these diverse assembly histories, it is essential to physically decompose galaxies into their fundamental structural components. While traditional photometric bulge-disc decomposition provides valuable morphological insights, it inherently suffers from degeneracies \citep[e.g.,][]{gadotti2009structural}. Overlapping structures, projection effects, and the lack of kinematic information mean that photometry alone cannot robustly distinguish between a pressure-supported classical bulge and a rotationally supported pseudo-bulge, failing to fully capture the true dynamical state of the underlying stellar populations.

    The advent of Integral Field Spectroscopy (IFS) provides stellar kinematical data across the two-dimensional sky plane of galaxies. By further applying dynamical methods, such as the orbit-superposition Schwarzschild model \citep{schwarzschild1979numerical, Rix1997ApJ...488..702R,van2008recovery, vdB2008MNRAS.385..647V, Gebhardt2000AJ....119.1157G, Eugene2013MNRAS.434.3174V}  (including DYNAMITE \citep[DYnamics, Age and Metallicity Indicators Tracing Evolution,][]{jethwa2020dynamite, thater2022testing}) and the Made-to-Measure model \citep{BlaaDaz2018MNRAS.481.3210B,Long2012MNRAS.421.2580L} to fit the stellar kinematical data, we can uncover the internal stellar orbit distribution of a galaxy \citep[e.g.,][]{vdB2008MNRAS.385..647V, zhu2018stellar, zocchi2023dynamical}. This method has been widely applied to large samples of galaxies from large IFS surveys across the Hubble sequence \citep{zhu2018stellar, santucci2022sami, Jin2020MNRAS.491.1690J}. A population-orbit superposition method has been further developed by tagging the orbits with stellar populations \citep{zhu2022b, zhu2022a,Jin2024A&A...681A..95J, Ding2023A&A...672A..84D}. Based on the stellar orbit distribution obtained from the dynamical models, we can dynamically decompose a galaxy into different components in a flexible and physically motivated way. In earlier work \citep{zhu2018stellar}, with the stellar orbit distribution of 300 CALIFA galaxies obtained, we dynamically decomposed each galaxy into dynamically cold, warm, hot, and counter rotating components and quantified the variation in luminosity fraction of each component as a function of galaxy stellar mass.

    Such dynamical decomposition leads to the direct comparison of galaxy structures from observations and simulations.
    For instance, \citet{xu2019} compared the stellar orbit fractions of TNG100 galaxies with those inferred from CALIFA \citep{zhu2018stellar} and found that the simulation broadly reproduces the observed mass dependence of the orbital components. Building upon these foundations, \citet{zhang2025diverse} recently expanded this analysis by comparing CALIFA observations with the high-resolution IllustrisTNG simulation suite TNG50 \citep{pillepich2019tng50, nelson2019tng50}. These works verified the broad success of the simulation in reproducing the internal structures of real galaxies. The stellar orbit distribution bridges the observations and simulations; the physical origins of different components can thus be further traced in the simulations.
    \citet{zhang2025diverse} highlighted that there is a mass-dependent divergence in the evolutionary paths of the dynamically hot orbital component. In high-mass galaxies ($M_{\star} \gtrsim 10^{10.5} M_{\odot}$), the simulations predict that the hot component is primarily composed of stars that were ``born cold'' in early, thin discs; these stars were subsequently heated into dispersion-dominated orbits through major mergers or persistent secular processes \citep{quinn1993heating, brook2004creation, grand2016vertical, sotillo2022merger, lagos2022diverse}. Conversely, in low-mass galaxies, the hot orbits are mainly populated by stars born hot. If such low-mass systems avoid subsequent major mergers, their hot component may preserve information about early formation.
    

    However, testing these mass-dependent evolutionary scenarios observationally remains difficult because the achievable physical resolution of the observational data varies strongly across integral-field surveys. The CALIFA data used by \citet{zhu2018stellar} have a typical point spread function (PSF) full width at half maximum (FWHM) of $\sim2.5\arcsec$, corresponding to sub-kpc to kpc-scale physical resolution across the survey redshift range. There might be relatively large uncertainty in the stellar orbit distribution of each single galaxy, especially for those with $M_{\star}\lesssim 10^{10}\,M_{\odot}$.
    MUSE provides data with significantly higher quality, especially with substantially better angular resolution, with an ideal seeing FWHM of $\sim0.4\arcsec$. For the galaxies studied here, this translates into physical resolutions ranging from tens to a few hundred parsecs, depending on the distance. In this paper, we aim to test whether higher-spatial-resolution MUSE data yield orbital decompositions broadly consistent with the mass-dependent trends previously reported from CALIFA, and to examine how individual galaxies deviate from the statistical results.
    
    This paper is organised as follows. In \autoref{sec:data}, we describe the observational data, detailing the sample selection from the MUSE archive. \autoref{sec:schwarzschild} outlines our data analysis methodology, including the reduction procedures utilising the GIST pipeline and the implementation of the Schwarzschild orbit-superposition technique, in which the triaxial code of \citet{vdB2008MNRAS.385..647V} is used in its axisymmetric limit. In \autoref{sec:results}, we present the main results of our dynamical modelling, focusing on the quality of the kinematic fits and the quantitative orbital decomposition of the sample. \autoref{sec:discussion} discusses the physical implications of these findings. Finally, we summarise our conclusions in \autoref{sec:conclusions}.


\section{Data}
\label{sec:data}
\subsection{MUSE archive data and initial sample selection}
\label{subsec:muse}
    

    The MUSE archive serves as a comprehensive repository for high-resolution integral field spectroscopic data obtained with the Very Large Telescope (VLT) \citep{bacon2010muse, bacon2017muse}. All data used here were taken in Wide Field Mode with the nominal setting, which delivers a $1\arcmin\times1\arcmin$ field of view sampled by $0.2\arcsec$ spaxels and covers 4750--9350~\AA\ at a constant $1.25$~\AA\,pixel$^{-1}$, with a line-spread function of FWHM $\simeq2.9$~\AA\ in the blue and $\simeq2.5$~\AA\ in the red \citep{bacon2017muse}. The galaxy-specific PSF was derived from the VLT slow-guiding-system (SGS) header. These header estimates span $0.45$--$1.90\arcsec$, with a median of $0.72\arcsec$.
    
    In this study, we utilised the archive to identify extragalactic targets containing three-dimensional (3D) data cubes. These cubes provide a rich dataset consisting of two spatial axes and one spectral axis, enabling spatially resolved analysis of galaxy kinematics. Our initial search focused on identifying all galaxy-type objects with available post-processed data products, yielding a total of 208 data cubes. We visually inspected these for overall image quality, signal-to-noise ratio (S/N), and the absence of significant instrumental artefacts. Since the primary objective of this project is to study the internal structures of disc systems, we restricted our selection to spiral or disc galaxies in the local Universe. We therefore retained 53 galaxies for initial analysis. A significant portion of our high-quality data is from the MUSE Atlas of Discs (MAD) survey \citep{erroz2019muse}.

\subsection{Deriving stellar kinematics using the GIST pipeline}

    \begin{figure*}
        \centering
        \includegraphics[width=\textwidth]{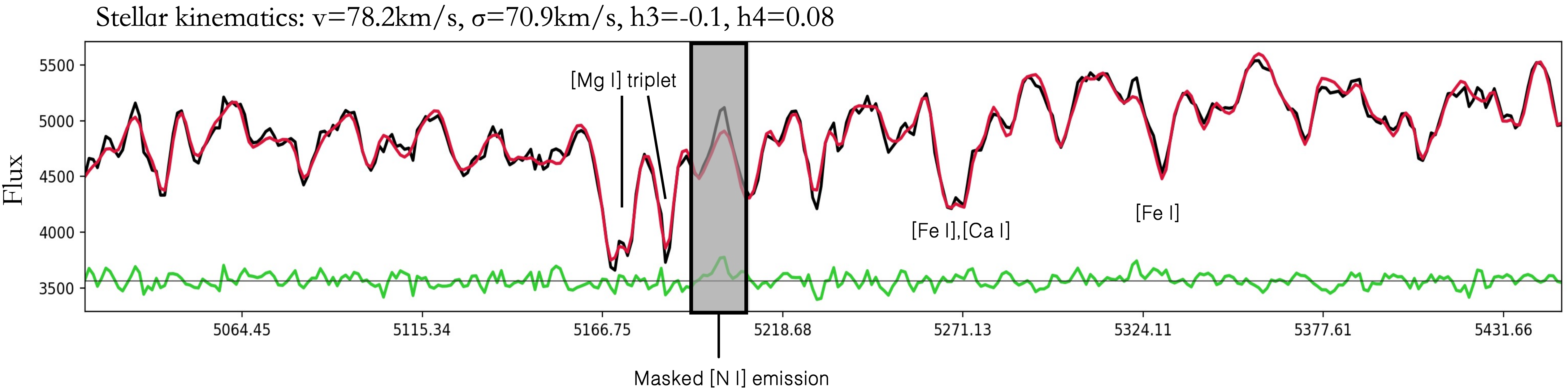}
           \caption{Illustration of the pPXF fit of a single spectrum within the optimised 5020--5450~\AA\ wavelength window. The black line represents the observed MUSE spectrum, while the red line indicates the best-fit stellar template. Residuals are shown in green at the bottom of the panel. The grey shaded region at $\sim$5200~\AA\ denotes the masked [N I] emission line to prevent contamination of the stellar kinematics. The primary absorption lines that provide kinematic information are also marked. The spectrum is shown in the rest frame.}
        \label{fig:ppxf}
    \end{figure*}

    \begin{figure*}[ht!]
        \centering
        \includegraphics[width=\textwidth]{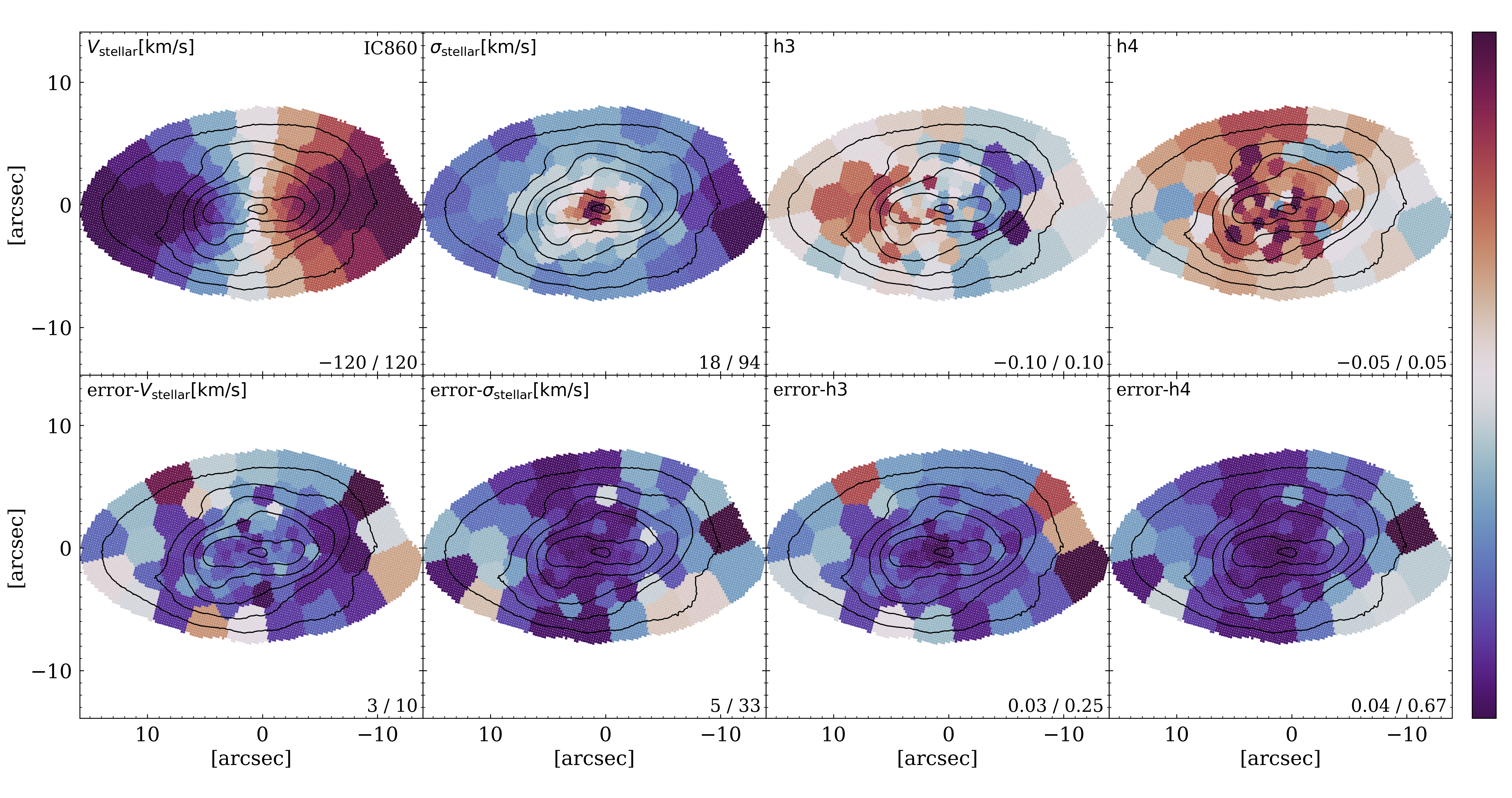}
           \caption{Spatially resolved stellar kinematics of IC\,860 extracted using the GIST pipeline. The upper panels show mean stellar velocity ($V_{\mathrm{stellar}}$), velocity dispersion ($\sigma_{\mathrm{stellar}}$), as well as the higher-order Gauss--Hermite coefficients $h_3$ and $h_4$, from left to right. The lower panels show the corresponding $1\sigma$ errors.}
        \label{fig:kinematics_example}
    \end{figure*}

    To extract the stellar kinematics, we utilise the Galaxy IFU Spectroscopy Tool (GIST) pipeline \citep{bittner2019gist}. GIST is a modular, parallelised framework designed for the analysis of IFS data, providing a workflow for data preparation, spatial binning, and physical property extraction, derived from the original Penalised PiXel-Fitting (pPXF) method \citep{cappellari2004parametric, cappellari2017improving}.
    
    We perform spatial binning using the Voronoi tessellation method to ensure reliable kinematic extraction. For every spaxel GIST computes the ratio of the median flux to the median noise, the latter taken from the pipeline \texttt{STAT} variance extension, over the same 5020--5450~\AA\ window used for the kinematic fit and not over the full MUSE range; the S/N is therefore a per-spaxel, per-spectral-pixel quantity. Whole spaxels with $\mathrm{S/N}<5$ are discarded. The rest are combined with the Voronoi algorithm of \citet{cappellari2003adaptive} to a nominal target $\mathrm{S/N}=100$ per bin; NGC\,3521 and NGC\,1385 instead use targets of 200 and 300, respectively. The final maps contain between 65 and 2807 bins per galaxy (median 363).
    
   We adopt an optimised wavelength interval of 5020--5450~\AA\ (\autoref{fig:ppxf}) to derive the stellar kinematics, and fit the spectra using the MILES stellar library \citep{sanchez2006medium, falcon2011updated}. This choice is driven by three main factors. First, in our gas-rich, star-forming discs the nebular emission becomes strong and spatially variable at wavelengths bluer than 5020~\AA\ (H$\beta$ and [O\,{\sc iii}]$\lambda\lambda4959,5007$), and subtracting it can introduce bin-to-bin systematics in $h_3$ and $h_4$. Second, while individual features (e.g. [O\,{\sc i}]$\lambda5577$) can be masked, at $\gtrsim 5500$~\AA\ the spectra include Na\,D interstellar absorption and the [N\,{\sc ii}]+H$\alpha$ complex, and beyond 7000~\AA\ these archival data show numerous OH sky-line residuals \citep{weilbacher2020data}. Third, we also measured the kinematics in the calcium-triplet region but found it less appropriate. Although the MUSE instrumental dispersion decreases to the red, from $73.7$~km\,s$^{-1}$ at 5020~\AA\ to $37.6$~km\,s$^{-1}$ at 8600~\AA, and the calcium-triplet library adopted for those fits (FWHM $=1.50$~\AA) remains narrower than the MUSE line-spread function there, the red-side fits nonetheless yielded more weakly constrained $h_3$ and $h_4$ and were impacted by OH residuals. Moreover, the usable calcium-triplet window ($8480$--$8680$~\AA) is only about half as wide as our fiducial range and is dominated by just the three Ca lines, whereas 5020--5450~\AA\ includes the Mg\,$b$ triplet plus neighbouring Fe and Ca blends that carry most of the higher-moment constraints. We compared this interval with near-infrared and broad optical ranges and found that it performs best. We further mask 5192--5204~\AA\ to exclude contamination from [N I] emission.
    
    The spectra are logarithmically rebinned to a constant $70$~km\,s$^{-1}$ per pixel, while the templates are sampled twice as finely. The MILES templates are broadened from their native FWHM of $2.51$~\AA\ to the MUSE line-spread function at each wavelength before the fit. We fit four Gauss--Hermite moments with additive and multiplicative polynomials of degree $-1$ and $0$, respectively, and retain the pPXF default penalty, $\mathrm{bias}=0.7\sqrt{500/N_{\rm pix}}$. The continuum shape is instead described by a fitted dust-reddening curve, which pPXF requires to be used in place of multiplicative polynomials.
    We verified that additive polynomial does not affect the recovered kinematics. We re-extracted every bin of six galaxies spanning the sample's mass range --- NGC\,3521 and NGC\,4941, which show a central rise in dispersion, together with the low-dispersion systems NGC\,1483, NGC\,4592, NGC\,4980 and NGC\,1042 --- using an additive polynomial of degree one and leaving every other setting unchanged. Across the 3378 bins the median absolute change in dispersion is $0.4$--$2.9$~km\,s$^{-1}$, between $0.04$ and $0.36$ times the formal pPXF uncertainty, and only 22 bins ($0.7$ per cent) move by more than their own error bar. The shape of the dispersion profile is preserved: the ratio of the median $\sigma$ within $0.25\,R_{\mathrm{e}}$ to that between $0.5$ and $1\,R_{\mathrm{e}}$ changes from $1.045$ to $1.056$ in NGC\,3521 and from $1.462$ to $1.465$ in NGC\,4941, while the four low-mass galaxies remain featureless, with ratios changing by less than $0.02$. The distinction between the centrally peaked dispersion profiles of the massive galaxies and the flat profiles of the low-mass ones is therefore a property of the data rather than of the polynomial degree.
    The kinematic information is predominantly provided by the stellar continuum and well-defined absorption features \citep{vazdekis2010evolutionary}, as marked in \autoref{fig:ppxf}, chiefly the Mg\,$b$ triplet at $\sim$5170~\AA, the Fe\,{\sc i} lines at 5270 and 5335~\AA, and the Ca\,{\sc i} lines at 5265.6 and 5270.3~\AA.

    An example of the final kinematic maps for IC\,860 is shown in \autoref{fig:kinematics_example}. The high sensitivity of the selected spectral indices ensures robust measurement of the stellar kinematics, including the mean velocity, velocity dispersion, and the higher-order Gauss--Hermite coefficients $h_3$ and $h_4$, minimising systematic errors while preserving the high spatial resolution of MUSE. A significant fraction of our spatial bins have velocity dispersions at or below the instrumental dispersion within the fitting window. In this regime, $h_3$ and $h_4$ are poorly constrained, which manifests as large uncertainties. Estimating $h_3$ and $h_4$ at dispersions close to the instrumental resolution—while propagating the uncertainties in this manner—is standard practice in integral-field spectroscopy \citep{falcon2006sauron, bittner2020inside}. In the central regions of some galaxies, the dispersion is higher, and $h_3$ and $h_4$ are therefore better constrained with smaller errors. We retain $h_3$ and $h_4$ for the subsequent analysis to preserve reliable constraints in the inner parts, and even values consistent with zero but with large uncertainties in the outer regions still provide useful information to the model. The observational uncertainties will naturally propagate into our model results.

  \begin{figure}
        \centering
        \includegraphics[width=\columnwidth]{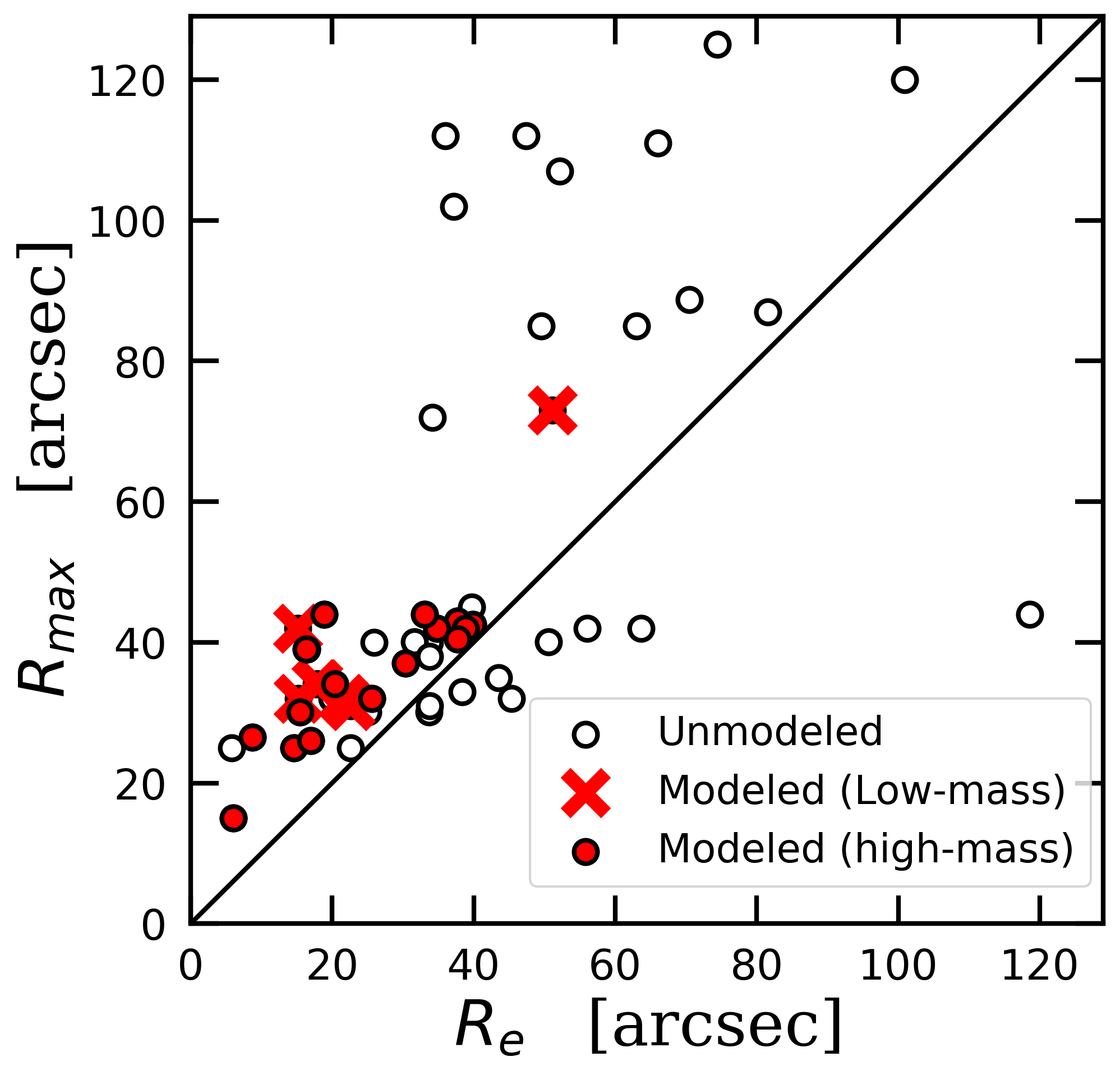}
        \caption{Distribution of $R_{\mathrm{max}}$ relative to $R_{\mathrm{e}}$ for our initial sample of 53 galaxies from the MUSE archive. The solid black line denotes $R_{\mathrm{max}} = R_{\mathrm{e}}$.         
        Red symbols mark the galaxies selected for Schwarzschild modelling: red crosses have $M_{\star}<10^{10}\,M_{\odot}$ and red filled circles have $M_{\star}>10^{10}\,M_{\odot}$. Open black circles are galaxies that pass the initial quality inspection but are not modelled. We only select those galaxies with data coverage extending to at least $1\,R_{\mathrm{e}}$ ($R_{\mathrm{max}} > R_{\mathrm{e}}$).}
        \label{fig:coverage}
    \end{figure}

    \begin{figure}
        \centering
        \includegraphics[width=\columnwidth]{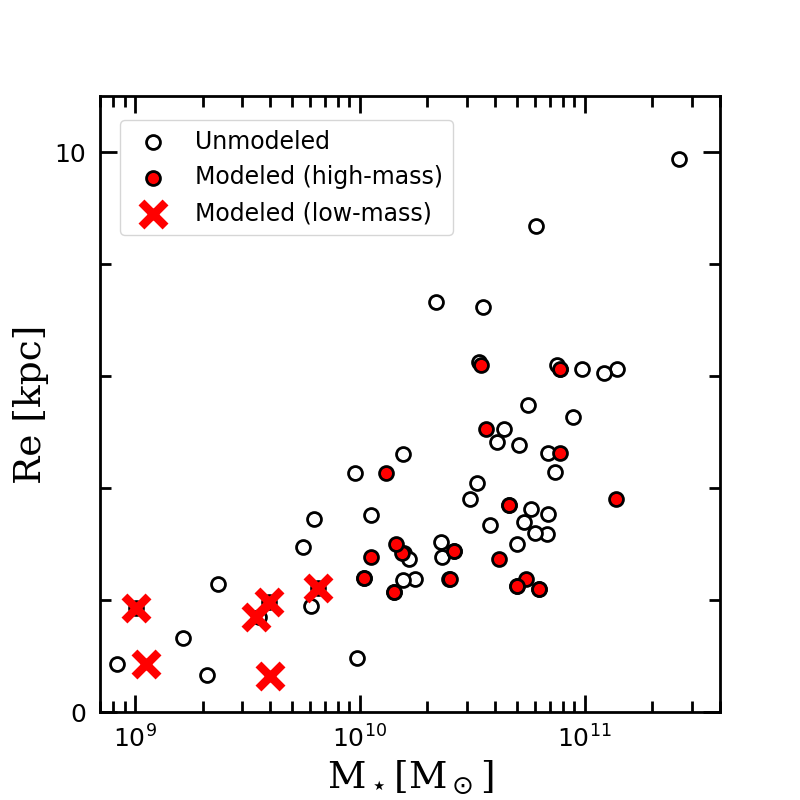}
        \caption{Stellar mass ($M_{\star}$) versus $R_{\mathrm{e}}$, with the same symbols as in \autoref{fig:coverage}.}
        \label{fig:mass_size}
    \end{figure}

    \begin{figure}
        \centering
        \includegraphics[width=\columnwidth]{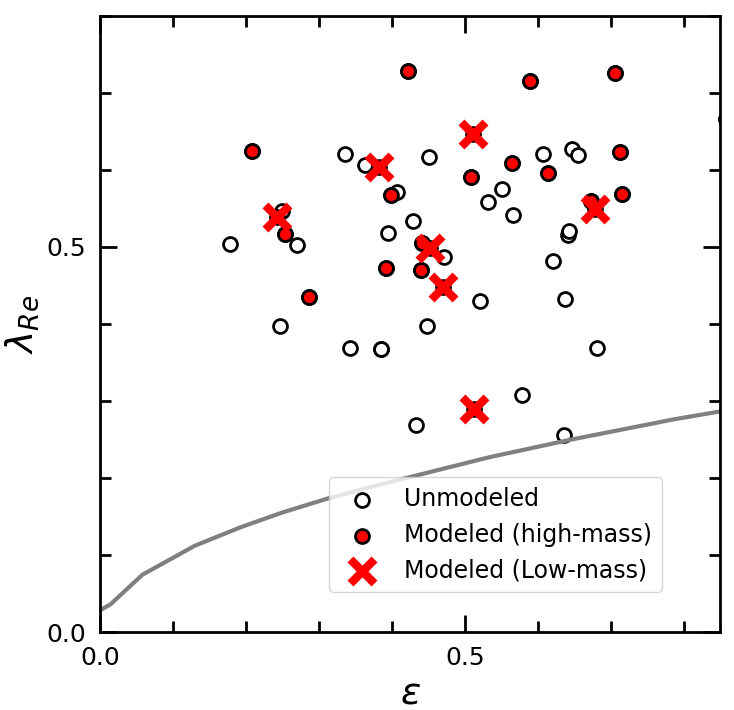}
        \caption{Ellipticity ($\epsilon$) versus the apparent stellar angular momentum within 1 ${R_{\mathrm{e}}}$ ($\lambda_{R_{\mathrm{e}}}$), with the same symbols as in \autoref{fig:coverage}. The grey curve demarcates the division between Slow and Fast rotators as established by \citet{emsellem2011atlas3d}.}
        \label{fig:epsilon_lambda}
    \end{figure}
    \subsection{Final sample for this work}
    
    A single MUSE pointing provides a typical maximum radial data coverage ($R_{\mathrm{max}}$) of approximately $30\arcsec$. Consequently, for some nearby or spatially extended galaxies in our initial sample, $R_{\mathrm{e}}$ exceeds the instrumental field of view. \autoref{fig:coverage} illustrates the distribution of radial coverage relative to $R_{\mathrm{e}}$ for all targets. Because our orbital decomposition and dynamical analysis require robust kinematic constraints out to at least $R_{\mathrm{e}}$, we exclude any system where $R_{\mathrm{max}} < R_{\mathrm{e}}$. Applying this spatial coverage criterion removes nine of the original 53 candidate galaxies. 
    
    \label{subsec:sample}
    Three kinds of system are excluded at this stage. First, galaxies close to face-on, whose
    line-of-sight velocity field carries too little information to constrain the orbit distribution.
    Second, galaxies with a strong bar, which violates the axisymmetry our models assume; we note that
    a strong bar is no longer an obstacle to orbit-superposition modelling in principle, since
    \citet{tahmasebzadeh2022barred} have extended the triaxial Schwarzschild method to barred systems,
    but applying that machinery is beyond the framework adopted here. Third, galaxies whose stellar
    kinematic maps are irregular or broken into fragments, which in practice arises where dust
    obscuration or strong, spatially concentrated star formation disrupts the stellar continuum.
    Examples of all three are shown in \autoref{fig:rejected}.

    Galaxies with a weak bar are retained. So are dust-rich galaxies, provided their stellar
    kinematic maps remain smooth and regular over the modelled field: the operative criterion is the
    state of the kinematics rather than the presence of dust as such. Four such galaxies are in the
    final sample, and they are marked $^{b}$ or $^{d}$ in \autoref{tab:best_model_summary}. 
    
    Our final sample for Schwarzschild modelling is denoted by the red symbols in \autoref{fig:coverage}. We include six galaxies with $M_{\star}<10^{10}\,M_{\odot}$ and 18 galaxies with $M_{\star}>10^{10}\,M_{\odot}$. A few of the galaxies were observed with multiple MUSE pointings, so that their data coverage $R_{\mathrm{max}}$ extends beyond the radius of a single pointing in \autoref{fig:coverage}.
    Because this sample is archive-selected and filtered for data quality and modelling suitability, it should not be interpreted as a statistically representative sample of local late-type galaxies.
    The redshifts span $0.00241\leq z\leq0.01920$ for the 53-galaxy initial sample and $0.00266\leq z\leq0.01920$ for the 24 modelled galaxies (median $z=0.00521$); individual catalogue distances are listed in \autoref{tab:basic_info}.
    
    The global structural properties of our final sample, specifically the relationship between stellar mass ($M_{\star}$) and $R_{\mathrm{e}}$, are presented in \autoref{fig:mass_size}.  Here we use stellar masses and effective radii from the catalogues compiled from the NASA/IPAC Extragalactic Database and \citet{erroz2019muse}. The stellar mass is compiled from a few published catalogues: MAD \citep{erroz2019muse} for 24 galaxies, z0MGS \citep{leroy2019z0mgs} for 18, S$^4$G \citep{sheth2010s4g} for nine, AMUSING++ \citep{lopezcoba2020amusing} for one, and the SDSS MPA-JHU catalogue for one. The value adopted for each galaxy and its source are listed in \autoref{tab:basic_info}, and the provenance of each catalogue is described in \autoref{sec:appendix_a}. 
    
    To further characterise the kinematical properties of these systems, we quantify the stellar angular momentum within $R_{\mathrm{e}}$, $\lambda_{R_{\mathrm{e}}}$, following the definition by \citet{emsellem2011atlas3d}. We calculate the apparent stellar angular momentum using the flux ($F_j$), distance from the centre ($R_j$), stellar velocity ($V_j$), and velocity dispersion ($\sigma_j$) of each pixel $j$ as:
    \begin{equation}
        \lambda_{\mathrm{R}} = \frac{\sum_{j}^{N} F_j R_j |V_j|}{\sum_{j}^{N} F_j R_j (V_j^2 + \sigma_j^2)^{\frac{1}{2}}} .
    \end{equation}
    The relationship between $\lambda_{R_{\mathrm{e}}}$ and the observed ellipticity ($\epsilon$) is shown in \autoref{fig:epsilon_lambda}. The ellipticity $\epsilon=1-b/a$ and photometric position angle were measured from the MUSE reconstructed white-light image with the \texttt{find\_galaxy} routine of the \textsc{MgeFit} package, which derives the centre, ellipticity and position angle from the first and second moments of the surface brightness of the brightest image pixels \citep{cappellari2002efficient}. All these galaxies are fast rotators, and the six galaxies with $M_{\star}<10^{10}\,M_{\odot}$ tend to have slightly lower $\lambda_{R_{\mathrm{e}}}$.

    The stellar mass models use archival imaging assembled galaxy by galaxy: 23 galaxies have \textit{HST} data from WFPC2, ACS/WFC, WFC3/UVIS, WFC3/IR or NICMOS, while ESO\,339-G011 uses a MUSE white-light image. The instruments, pixel scales, PSFs, MGE fitting and radial coverage are documented in \autoref{sec:appendix_mge}, which also shows the observed and model isophotes for all 24 galaxies.



\section{Schwarzschild orbit-superposition method}
\label{sec:schwarzschild}
    \begin{figure}
        \centering
        \includegraphics[width=\columnwidth]{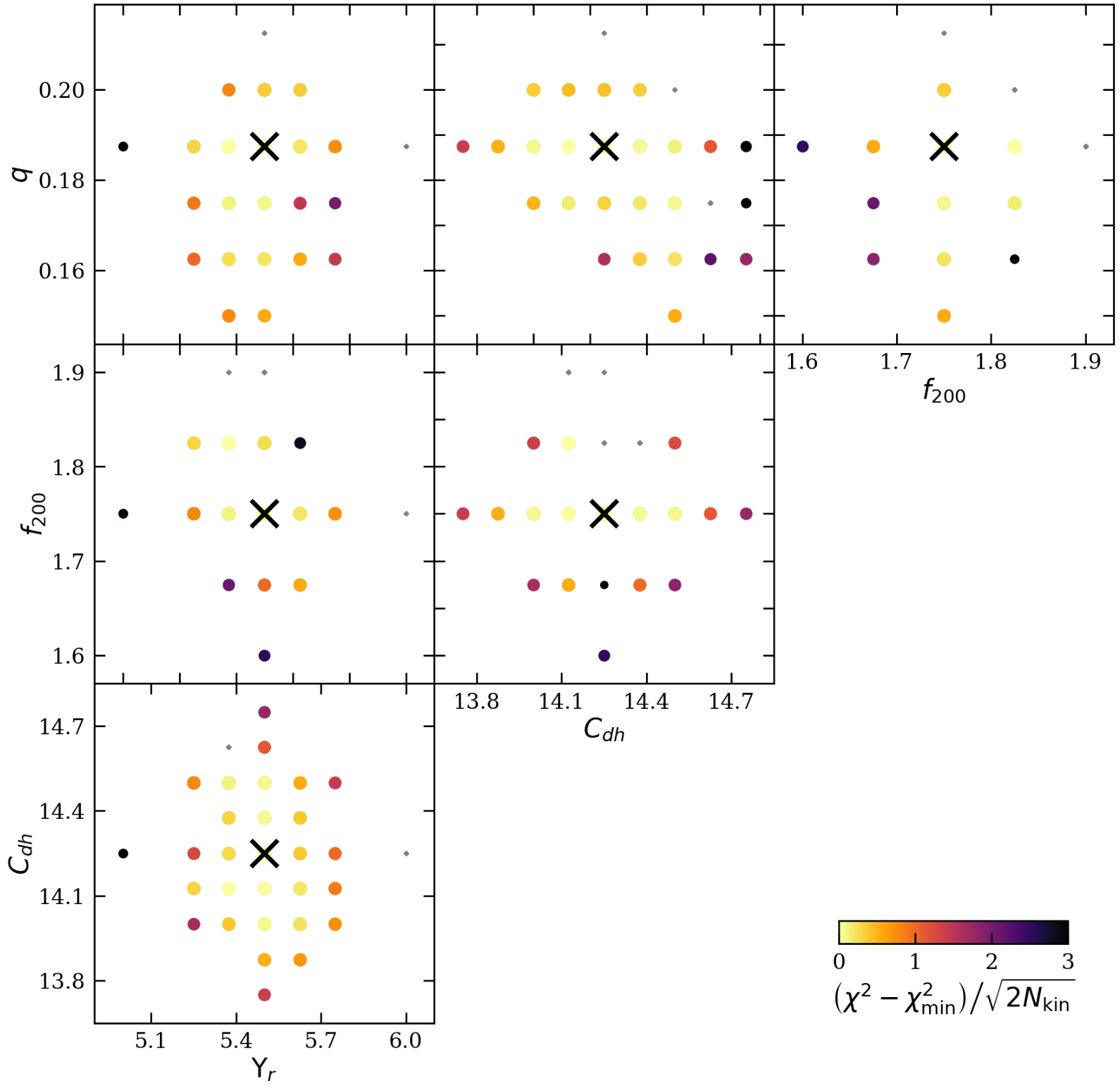}
        \caption{The parameter space spanned by our grid and the best-fitting Schwarzschild model for IC\,860. The dots lying within the $3\sigma$ confidence region are coloured by $(\chi^2 - \chi^2_{\mathrm{min}})/\sqrt{2N_{\mathrm{kin}}}$, with $N_{\mathrm{kin}} = n_{\mathrm{GH}}\times N_{\mathrm{obs}}$. Lighter colours indicate smaller values and therefore better fits. The grey dots represent models that fall outside the $3\sigma$ confidence region. The global best-fit model is marked with a black cross. The four model parameters explored are: $q$, $f_{200}$, $C_{dh}$, $\Upsilon_{\mathrm{r}}$. Note that the $f_{200}$ axis is logarithmic.}
        \label{fig:corner}
    \end{figure}

    To recover the intrinsic three-dimensional structure and mass distribution, we employ the Schwarzschild orbit-superposition method. We utilise the public DYNAMITE code \citep{jethwa2020dynamite, thater2022testing, zocchi2023dynamical}, a modern Python-based implementation of the triaxial Schwarzschild code of \citet{vdB2008MNRAS.385..647V}, itself building on \citet{schwarzschild1979numerical} and \citet{cappellari2002efficient}, optimised for high-resolution IFS data. We follow the standard way of constructing the model as described in \citet{Zhu2018MNRAS.473.3000Z} and \citet{jethwa2020dynamite}; here we briefly describe the major steps.

    \begin{figure*}[ht!]
        \centering
        \includegraphics[width=\hsize]{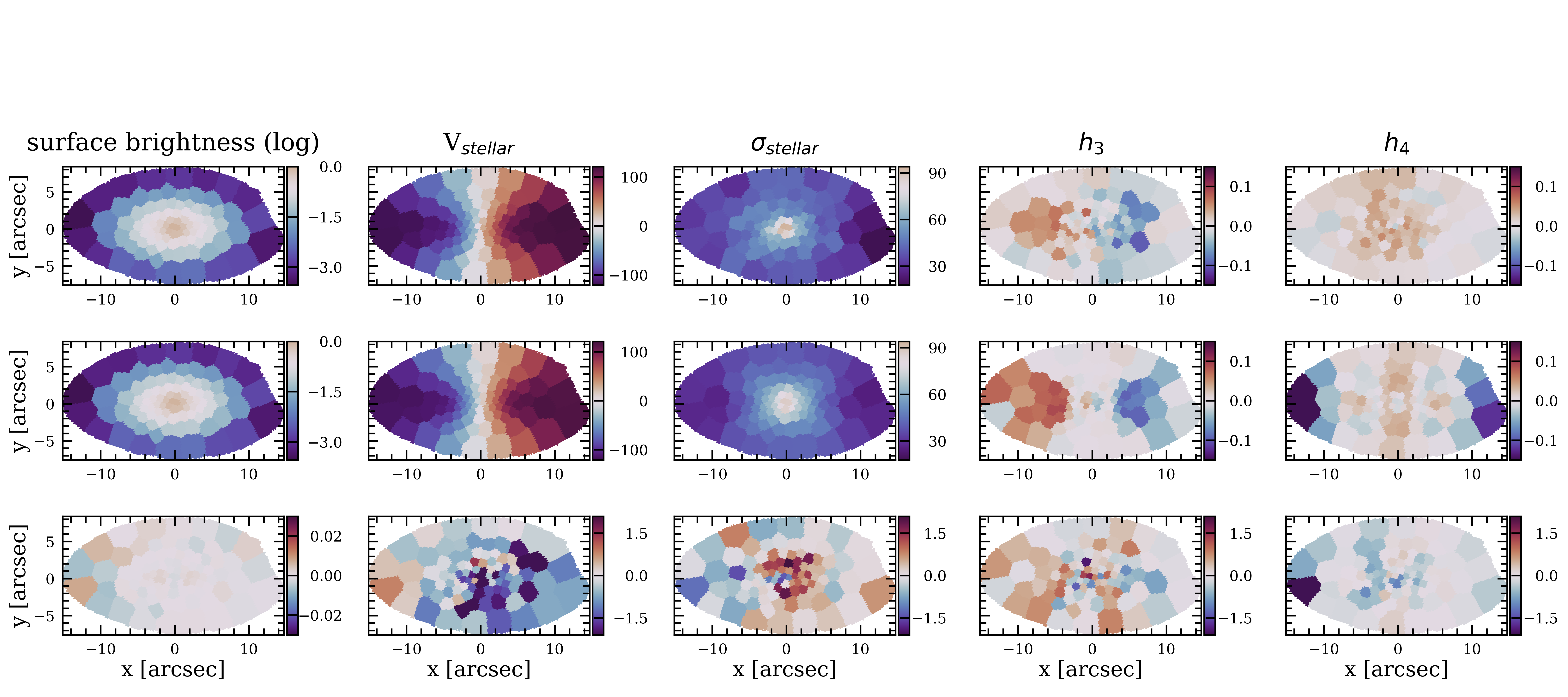}
        \caption{Comparison between observed MUSE kinematics and the best-fit Schwarzschild model for IC\,860. The three rows display the observed data (top), the model (middle), and normalised residuals (bottom). Columns from left to right are SB, $V_{\mathrm{stellar}}$, $\sigma_{\mathrm{stellar}}$, $h_3$ and $h_4$. Similar figures for the other galaxies are shown in \autoref{sec:appendix_kinmaps}.}
        \label{fig:model_example}
    \end{figure*}

\subsection{Gravitational potential}

    We construct a model of the gravitational potential by combining the contributions of stars, dark matter (DM), and a central supermassive black hole (BH). 
    For the stellar mass, we take the photometric image and parameterise the stellar surface brightness (SB) distributions using the Multi-Gaussian Expansion \citep[MGE;][]{emsellem1994mge,cappellari2002efficient} method, as implemented in the \texttt{mge\_fit\_sectors} routine of the \textsc{MgeFit} package. The photometry is assembled galaxy by galaxy from archival imaging, predominantly \textit{HST} (WFPC2, ACS/WFC, WFC3/UVIS, WFC3/IR and NICMOS), with a MUSE white-light image for ESO\,339-G011. The frames, their sampling, the PSF adopted for each deconvolution and the resulting fit quality are documented in \autoref{sec:appendix_mge}. 
    
    We de-project the 2D MGE to obtain the 3D luminosity density by assuming viewing angles of the galaxy. Given the properties of each Gaussian (flattening $q_{\rm obs}$, dispersion $\sigma_{\rm obs}$) on the 2D observational plane, the three viewing angles ($\theta$, $\psi$, $\phi$) are connected to the three parameters describing the intrinsic shape of the Gaussian ($q=c/a$, $p=b/a$, $u = \sigma_{\rm obs}/\sigma_{\rm intr}$), where $a$, $b$ and $c$ are the long, intermediate, and short axes, and $\sigma_{\rm intr}$ is the intrinsic dispersion. We take $q,p,u$ instead of the three viewing angles as free parameters in our model. Given that our final sample consists entirely of 24 disc and spiral galaxies that are close to axisymmetric, we fix $p = 0.999$ and $u = 0.9999$, and allow $q$ to be free. We further multiply a stellar mass-to-light ratio $\Upsilon_r$ by the luminosity density to obtain the stellar mass density distribution.

    We adopted a Navarro--Frenk--White \citep[NFW;][]{navarro1997universal} DM halo with two free parameters: the virial mass $M_{\rm 200}$ and concentration $C_{dh}$. The BH sphere of influence is not resolved by the MUSE data in our galaxies and will not affect our fitting to the data; we add the BH as a point with a fixed mass of $M_{\mathrm{BH}} = 10^6 M_{\odot}$ and a nominal scale length of $a = 0.001$. The typical influence radius of the central BH in our galaxies is $\lesssim 0.1$ arcsec, much smaller than the PSF of the MUSE datacube, thus not resolved by the kinematic data. For example, adopting the $M_{\mathrm{BH}}$--$M_{\star}$ relation of \citet{reines2015relations}, IC\,2560, the most massive galaxy of the sample, would host $M_{\mathrm{BH}}\simeq2\times10^{7}\,M_{\odot}$ and hence $r_{\mathrm{SOI}}=GM_{\mathrm{BH}}/\sigma_{\mathrm{e}}^{2}\simeq4$--$9$~pc, or $0.03$--$0.06\arcsec$ at 33.4~Mpc. Including a BH with fixed mass will thus not affect our results.

    We thus have four free hyperparameters in the gravitational potential: the intrinsic flattening $q$ relates to the inclination angle, the stellar mass-to-light ratio $\Upsilon_r$, the DM virial mass $M_{\rm 200}$ scaled by stellar mass $f_{\rm 200} =M_{\rm 200}/M_{\star}$, and concentration $C_{dh}$.  We adopt $H_0=73.0\ {\rm km\,s^{-1}\,Mpc^{-1}}$ and define $M_{200}$ within the radius whose mean enclosed density is 200 times the critical density evaluated at $z=0$.

\subsection{Orbit sampling and determination of orbit weights}

    For each set of hyper-parameters, we compute tens of thousands of orbits in the corresponding gravitational potential. 
    The orbit sampling and calculation follow what we described in \citet{vdB2008MNRAS.385..647V} and \citet{Zhu2018MNRAS.473.3000Z}. The active libraries all use $n_E=22$ energy shells and a dithering factor of three, with $(n_{I_2},n_{I_3})=(15,15)$. Each potential is sampled by two separate libraries, one of tube orbits and one of box orbits, so that these grids contain 267,300 integrated orbits per potential, grouped into 9900 orbit bundles whose weights are the free parameters of the linear fit. Every orbit is integrated for 200 orbital periods and sampled at 50,000 points in the meridional plane, with an integration accuracy of $10^{-5}$. The energy grid is logarithmically spaced in the radius of the circular orbit of the same energy, from $\log r=-1.0$ to $1.99$.
    The weights of the orbits in the resulting orbit library are then determined by fitting the orbit-superposition simultaneously to the intrinsic and projected luminosity density and stellar kinematic maps. The searches contain 4243 completed models, with a median of 160 per galaxy, run on 224 CPU cores for most galaxies and 112 for two. Together they required of order 1300 wall-hours, or approximately $2.7\times10^{5}$ CPU core-hours.

    \begin{table*}
    \centering
    \caption{Best-fit Dynamical Model Summary. Columns from left to right: (1) Galaxy identifier; (2) Total stellar mass derived from DYNAMITE models; (3) Intrinsic axial ratio ($q$); (4) Dark matter halo concentration parameter ($C_{dh}$); (5) Ratio of dark-matter halo virial mass to stellar mass, $f_{200}\equiv M_{200}/M_{\star}$; (6) Stellar mass-to-light ratio in the $r$ band ($\Upsilon_{\mathrm{r}}$); (7)--(10): Orbital luminosity fractions for the cold, warm, hot, and counter-rotating (CR) components, respectively. The quoted fraction uncertainties are the sample standard deviations across 100 kinematic Monte-Carlo realisations; the procedure is described in \autoref{subsec:decomp}. Galaxies marked $^{d}$ are dust-rich.}
    \label{tab:best_model_summary}
    \resizebox{\textwidth}{!}{%
    \begin{tabular}{lccccccccc}
    \hline\hline
    Galaxy & $M_{\star-\mathrm{DYNAMITE}}$ & $q$ & $C_{dh}$ & $f_{200}$ & $\Upsilon_{\mathrm{r}}$ & Cold & Warm & Hot & CR \\
     & ($M_{\odot}$) & & & & & & & & \\
    \hline
    ESO339-G011$^{d}$ & $6.0 \times 10^{10}$ & 0.050 & 13  & 40  & 7.0  & $0.32 \pm 0.081$ & $0.41 \pm 0.10$ & $0.26 \pm 0.045$ & $0.0090 \pm 0.016$ \\
    ESO498-G5   & $1.1 \times 10^{10}$ & 0.12  & 9.5 & 89  & 2.3  & $0.23 \pm 0.062$ & $0.52 \pm 0.080$ & $0.14 \pm 0.025$ & $0.11 \pm 0.041$ \\
    IC860       & $1.8 \times 10^{10}$ & 0.19  & 14  & 56  & 5.5  & $0.38 \pm 0.062$ & $0.44 \pm 0.082$ & $0.19 \pm 0.054$ & $0.00028 \pm 0.010$ \\
    IC2560$^{b}$      & $7.4 \times 10^{10}$ & 0.30  & 12  & 20  & 9.1  & $0.18 \pm 0.020$ & $0.57 \pm 0.036$ & $0.19 \pm 0.036$ & $0.067 \pm 0.0062$ \\
    NGC0289$^{b}$     & $4.6 \times 10^{10}$ & 0.19  & 8.8 & 110 & 5.3  & $0.43 \pm 0.021$ & $0.32 \pm 0.020$ & $0.19 \pm 0.012$ & $0.057 \pm 0.0084$ \\
    NGC0833     & $5.4 \times 10^{10}$ & 0.25  & 14  & 40  & 5.1  & $0.15 \pm 0.032$ & $0.52 \pm 0.038$ & $0.30 \pm 0.027$ & $0.024 \pm 0.012$ \\
    NGC1042     & $9.7 \times 10^{8}$  & 0.23  & 14  & 94  & 4.3  & $0.052 \pm 0.062$ & $0.39 \pm 0.17$ & $0.31 \pm 0.088$ & $0.24 \pm 0.086$ \\
    NGC1084     & $2.1 \times 10^{10}$ & 0.15  & 12  & 28  & 2.3  & $0.28 \pm 0.010$ & $0.50 \pm 0.012$ & $0.12 \pm 0.013$ & $0.10 \pm 0.015$ \\
    NGC1309     & $1.8 \times 10^{10}$ & 0.050 & 14  & 47  & 1.4  & $0.34 \pm 0.067$ & $0.44 \pm 0.082$ & $0.085 \pm 0.032$ & $0.13 \pm 0.040$ \\
    NGC1320     & $3.3 \times 10^{10}$ & 0.055 & 8.9 & 79  & 6.9  & $0.21 \pm 0.040$ & $0.45 \pm 0.043$ & $0.17 \pm 0.015$ & $0.18 \pm 0.018$ \\
    NGC1326$^{b}$     & $3.1 \times 10^{10}$ & 0.060 & 7.5 & 10  & 5.0  & $0.23 \pm 0.019$ & $0.46 \pm 0.022$ & $0.21 \pm 0.022$ & $0.10 \pm 0.019$ \\
    NGC1385$^{b}$     & $9.6 \times 10^{9}$  & 0.21  & 14  & 110 & 8.9  & $0.15 \pm 0.030$ & $0.48 \pm 0.041$ & $0.35 \pm 0.030$ & $0.020 \pm 0.017$ \\
    NGC1483$^{b}$     & $6.2 \times 10^{9}$  & 0.090 & 10  & 160 & 0.30 & $0.11 \pm 0.11$ & $0.60 \pm 0.16$ & $0.16 \pm 0.082$ & $0.12 \pm 0.058$ \\
    NGC3521$^{d}$     & $3.3 \times 10^{10}$ & 0.30  & 13  & 130 & 6.8  & $0.23 \pm 0.015$ & $0.40 \pm 0.014$ & $0.27 \pm 0.0095$ & $0.10 \pm 0.015$ \\
    NGC4030     & $5.7 \times 10^{10}$ & 0.11  & 13  & 150 & 2.8  & $0.44 \pm 0.015$ & $0.38 \pm 0.0097$ & $0.14 \pm 0.0063$ & $0.042 \pm 0.013$ \\
    NGC4418$^{d}$     & $1.7 \times 10^{10}$ & 0.030 & 9.4 & 10  & 7.5  & $0.26 \pm 0.041$ & $0.39 \pm 0.060$ & $0.28 \pm 0.039$ & $0.067 \pm 0.018$ \\
    NGC4592     & $4.7 \times 10^{8}$  & 0.21  & 12  & 32  & 0.30 & $0.25 \pm 0.21$ & $0.59 \pm 0.17$ & $0.15 \pm 0.17$ & $0.014 \pm 0.17$ \\
    NGC4790$^{b}$     & $3.7 \times 10^{9}$  & 0.15  & 11  & 100 & 0.60 & $0.27 \pm 0.16$ & $0.45 \pm 0.15$ & $0.17 \pm 0.10$ & $0.098 \pm 0.069$ \\
    NGC4941     & $2.1 \times 10^{10}$ & 0.30  & 11  & 220 & 7.0  & $0.14 \pm 0.015$ & $0.62 \pm 0.014$ & $0.12 \pm 0.019$ & $0.12 \pm 0.014$ \\
    NGC4980     & $7.8 \times 10^{8}$  & 0.26  & 9.3 & 320 & 0.35 & $0.23 \pm 0.22$ & $0.66 \pm 0.21$ & $0.092 \pm 0.19$ & $0.017 \pm 0.12$ \\
    NGC5806     & $4.4 \times 10^{10}$ & 0.12  & 7.5 & 10  & 2.9  & $0.17 \pm 0.018$ & $0.47 \pm 0.021$ & $0.26 \pm 0.022$ & $0.10 \pm 0.021$ \\
    NGC6754$^{b}$     & $4.1 \times 10^{10}$ & 0.12  & 8.3 & 150 & 1.4  & $0.39 \pm 0.056$ & $0.48 \pm 0.059$ & $0.10 \pm 0.016$ & $0.031 \pm 0.016$ \\
    NGC7162     & $1.7 \times 10^{10}$ & 0.16  & 13  & 24  & 2.5  & $0.44 \pm 0.089$ & $0.52 \pm 0.074$ & $0.022 \pm 0.025$ & $0.010 \pm 0.017$ \\
    NGC7771$^{b,d}$     & $6.0 \times 10^{10}$ & 0.15  & 15  & 320 & 6.2  & $0.23 \pm 0.020$ & $0.52 \pm 0.021$ & $0.12 \pm 0.016$ & $0.13 \pm 0.015$ \\
    \hline
    \end{tabular}}
    \tablenotes{b}{\textwidth}{Weak barred galaxies.}
    \tablenotes{d}{\textwidth}{Dust-rich galaxies.}
    \end{table*}

\subsection{Hyper-parameter grid searching and the best-fitting model}

    We calculate hundreds of models with a different collection of hyper-parameters. 
    The orbital weights are obtained with a non-negative least-squares solver, and each model is
    assessed with the kinematic $\chi^2$ evaluated directly on the Gauss--Hermite maps,
    \begin{equation}
        \chi^{2}_{\rm kin} \;=\; \sum_{c}\;\sum_{j=1}^{N_{\rm obs}}
        \left( \frac{F^{\rm obs}_{c,j}-F^{\rm mod}_{c,j}}{\delta F_{c,j}} \right)^{2},
        \label{eq:kinchi2}
    \end{equation}
    where $c$ runs over the $n_{\rm GH}=4$ fitted moments $\{V,\sigma,h_3,h_4\}$ and $j$ over the
    $N_{\rm obs}$ Voronoi bins, $F^{\rm obs}_{c,j}$ and $\delta F_{c,j}$ are the observed value and its
    pPXF uncertainty, and $F^{\rm mod}_{c,j}$ is the corresponding model quantity, obtained by fitting the
    same Gauss--Hermite parameterisation to the model line-of-sight velocity distribution in each bin so
    that model and data are compared like with like. No additional systematic term is added to
    $\delta F_{c,j}$ for these four moments. \autoref{eq:kinchi2} is the quantity minimised in the
    hyper-parameter search.

    Once we have explored the hyper-parameter space, we can select the best-fitting models with $\Delta\chi^2 \equiv \chi^2- \min(\chi^2) < \sqrt{2 N_{\rm kin}}$, following the scaling introduced by \citet{vdB2008MNRAS.385..647V}, where $N_{\rm kin}\equiv n_{\rm GH}\times N_{\rm obs}$, $n_{\rm GH}=4$ is the number of fitted stellar kinematic moments and $N_{\rm obs}$ is the number of Voronoi bins of each kinematic map. This is the same normalisation used for the colour scale of \autoref{fig:corner}, on which a model lying on the $1\sigma$ ($3\sigma$) boundary has $\Delta\chi^2/\sqrt{2N_{\rm kin}}=1$ (3). In \autoref{fig:corner}, we show the parameter space we explored for IC\,860. There are 22 models within the $1\sigma$ confidence level. All four hyper-parameters are generally well constrained, although there is relatively large uncertainty for the DM mass due to limited data coverage. In \autoref{tab:best_model_summary}, we summarise the parameters of the best-fit models for all 24 galaxies.

    The best-fitting model of IC\,860 is shown in \autoref{fig:model_example}. The model captures global rotation patterns and central dispersion structures. The higher-order moments, which are sensitive to the skewness and kurtosis of the line-of-sight velocity distribution (LOSVD), are well reproduced. The residuals are generally low and exhibit no strong pattern across the field of view.

\section{Results}
\label{sec:results}

\subsection{Orbital structure decomposition}
\label{subsec:decomp}

    \begin{figure}
        \centering
        \includegraphics[width=\columnwidth]{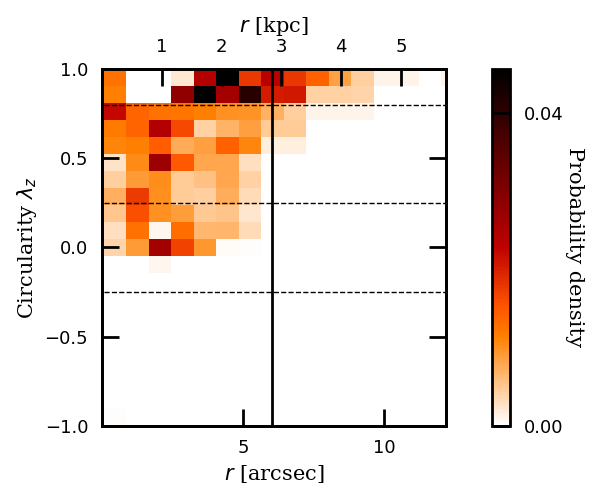}
        \caption{Distribution of orbital circularity ($\lambda_z$) as a function of radius $r$ (in arcsec) for IC\,860. $\lambda_z = 0.8, 0.25, -0.25$ are denoted with horizontal dashed lines. The vertical solid line indicates $R_{\mathrm{e}}$.}
        \label{fig:circularity}
    \end{figure}

    \begin{figure*}[ht!]
        \centering
        \includegraphics[width=\hsize]{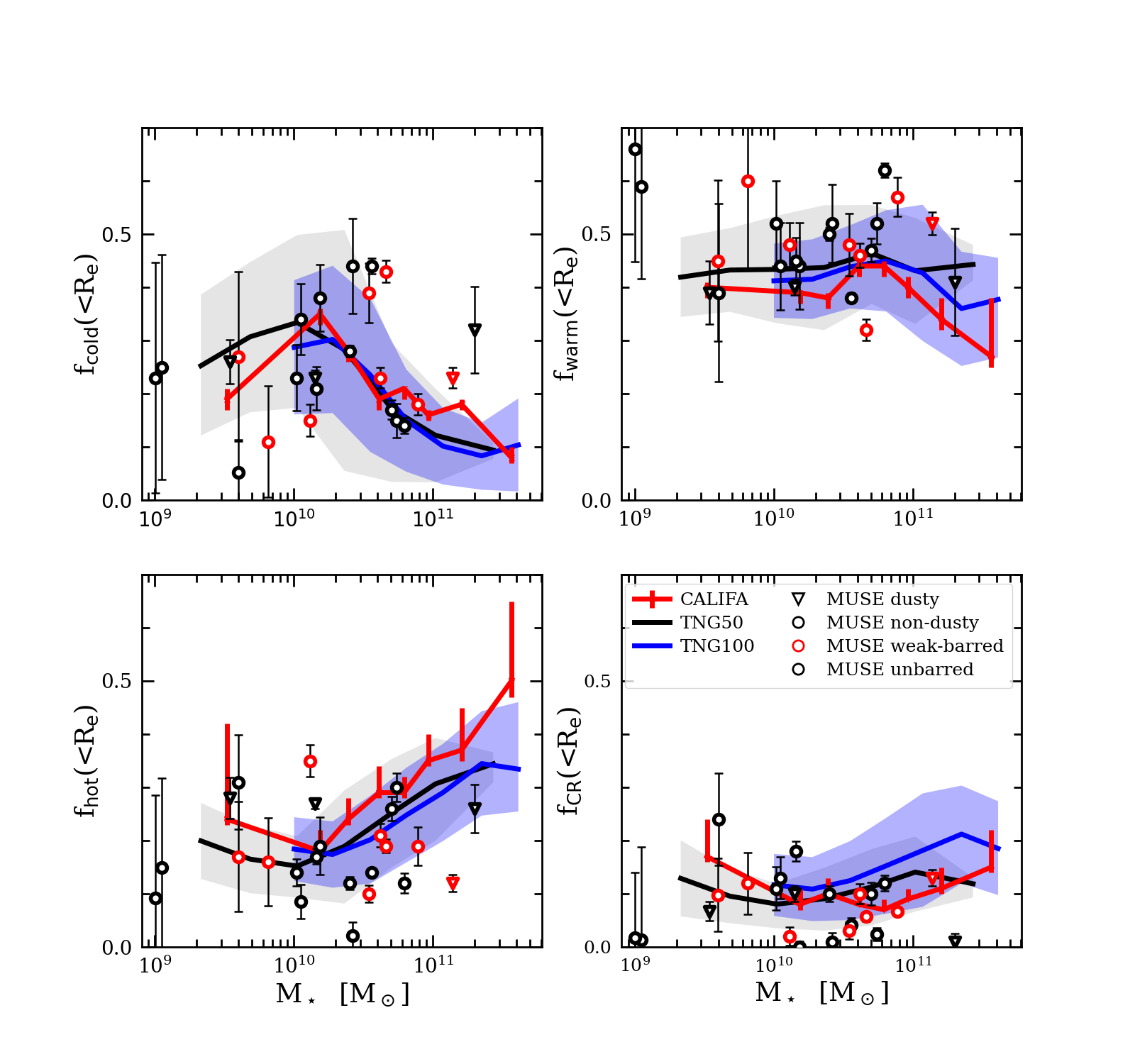}
        \caption{Luminosity fractions of orbital components within $R_{\mathrm{e}}$ as a function of stellar mass ($M_{\star}$, see \autoref{sec:appendix_a} for details). Panels show cold (top-left), warm (top-right), hot (bottom-left), and counter-rotating (bottom-right) orbits. Error bars show the standard deviation across 100 realisations of the kinematic Monte-Carlo procedure. Individual symbols show our MUSE sample: marker shape encodes dust content and edge colour encodes the presence of a weak bar, as labelled in the legend. Red lines show CALIFA median trends \citep{zhu2018stellar}; blue/black lines represent IllustrisTNG medians with shaded $1\sigma$ intervals \citep{zhang2025diverse, xu2019}.}
        \label{fig:decomposition}
    \end{figure*}

    A key feature of DYNAMITE is its capability to uncover the internal stellar orbit distribution. We then characterise each stellar orbit with two main properties \citep{zhu2018stellar}: the time-averaged radius, $r$, which represents the extent of the orbit, and the circularity, $\lambda_z = J_z/J_{\rm max} (E)$, which represents the angular momentum of the orbit around the minor $z$-axis, normalised by the maximum of a circular orbit with the same binding energy $E$.
    Whereas $|\lambda_z| \sim 1$ represents dynamically cold orbits dominated by regular rotation,
    $\lambda_z \sim 0$ represents dynamically hot orbits, which are dominated by radial random motions or long-axis tubes with regular rotation around the long axis.
    Negative values $\lambda_z<0$ refer to orbits that counter-rotate with respect to the net (prograde) rotation.
    \autoref{fig:circularity} presents the resulting stellar-orbit distributions of IC\,860 as probability-density distributions $p(r, \lambda_z)$ in the phase space of $r$ versus $\lambda_z$.
    Based on the stellar orbit distribution, we categorise orbits by circularity ($\lambda_z$): cold ($\lambda_z \ge 0.8$), warm ($0.25 < \lambda_z < 0.8$), hot ($|\lambda_z| \le 0.25$), and counter-rotating (CR) components ($\lambda_z < -0.25$) following \citet{zhu2018stellar}. We performed the analysis for all 24 galaxies and obtained their luminosity fractions of the four dynamical components.
    \autoref{fig:decomposition} shows that these galaxies are broadly consistent with the previously reported results from the CALIFA galaxies \citep{zhu2018stellar}, as well as the predictions from the IllustrisTNG simulations \citep{xu2019, zhang2025diverse}.

    We estimate the uncertainty that the kinematic measurement errors induce on the orbital fractions with a Monte-Carlo procedure. For each galaxy we generate 100 realisations of the kinematic maps by perturbing $V$, $\sigma$, $h_3$ and $h_4$ in every Voronoi bin by a Gaussian deviate with the standard deviation set to that bin's pPXF uncertainty, redrawing whenever a realisation returns a non-positive $\sigma$. Each realisation is then re-fitted with the gravitational potential and the orbit library held fixed at those of the best-fitting model, so that only the non-negative least-squares orbital weights are re-solved. The uncertainties quoted in \autoref{tab:best_model_summary} and shown in \autoref{fig:decomposition} are the standard
    deviations of the resulting orbital fractions across the 100 realisations.
    
    The galaxies with $M_{\star}\sim 2\times 10^{10}\,M_{\odot}$ have the highest luminosity fraction of the dynamically cold orbits. In the lower-mass galaxies, the luminosity fraction of cold orbits decreases, consistent with the trend exhibited in the CALIFA galaxies \citep{zhu2018stellar}; several of these galaxies have comparatively large warm and hot orbit fractions. Our two lowest-mass galaxies exhibit relatively low cold component fractions. Rather than being outliers, these systems lie on the low-mass extension of the CALIFA and TNG relations \citep{zhu2018stellar, xu2019, zhang2025diverse}. The distinction is between components: in $f_{\rm cold}$, $f_{\rm hot}$ and $f_{\rm CR}$ the two systems follow the low-mass extensions of the comparison relations, while their warm fractions are elevated, at 0.59 and 0.66. From $M_{\star} \gtrsim 2\times 10^{10}\,M_{\odot}$, we observe that the luminosity fraction of the cold component decreases, while the luminosity fraction of the hot component increases towards the high-mass end. The counter-rotating fraction remains low in most objects; there are no galaxies with obvious counter-rotating discs in our sample.

    \label{subsec:robustness}

    \begin{figure}[ht!]
        \centering
        \includegraphics[width=\hsize]{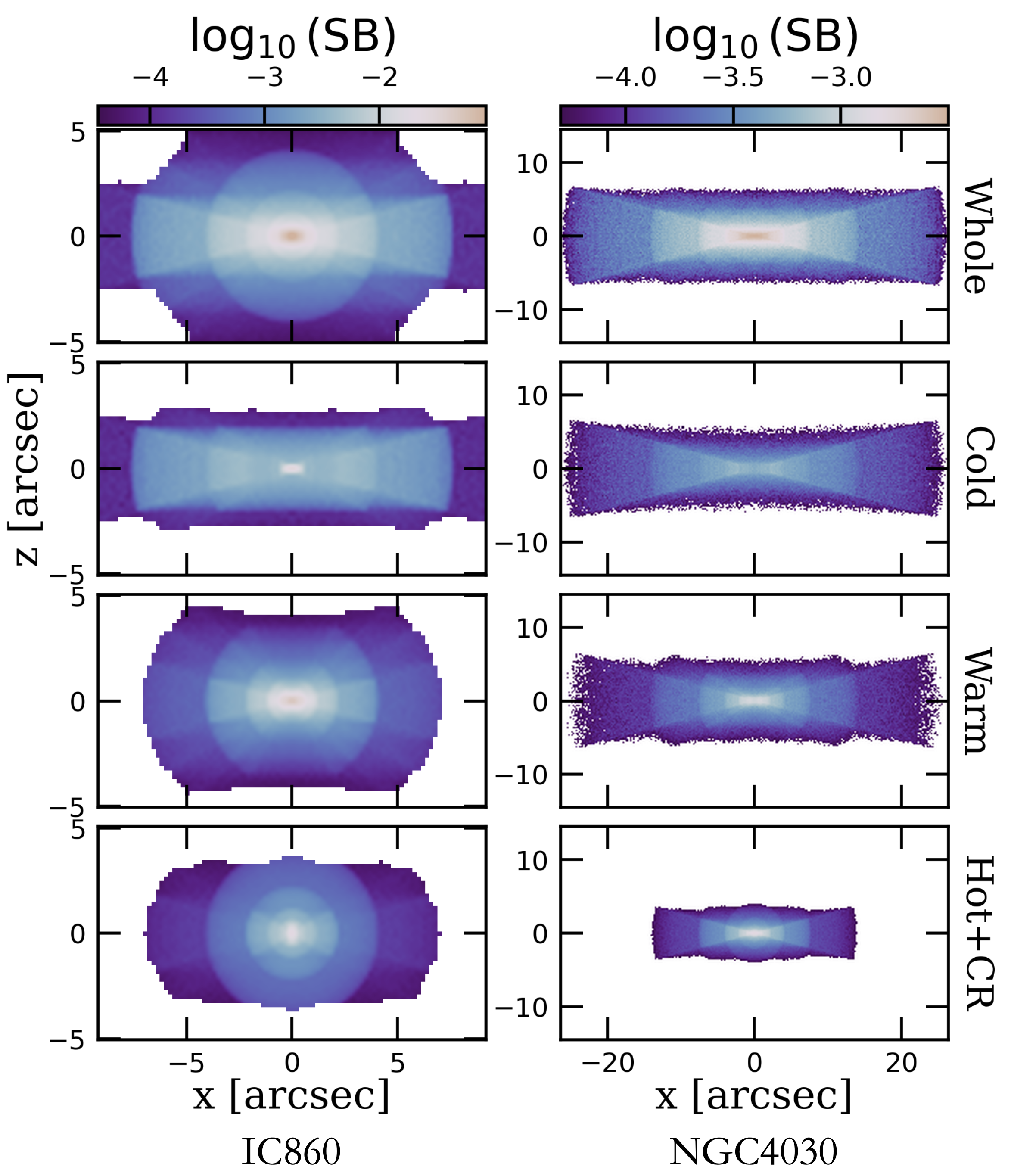}
        \caption{The edge-on surface brightness map of the whole galaxy, and of the cold, warm, and hot+CR components rebuilt by our model. We illustrate the morphologies for two galaxies: IC\,860 on the left and NGC\,4030 on the right.}
        \label{fig:morp_comp}
    \end{figure}

\subsection{The morphology of orbital components}
    \begin{figure*}[ht!]
        \centering
        \includegraphics[width=\hsize]{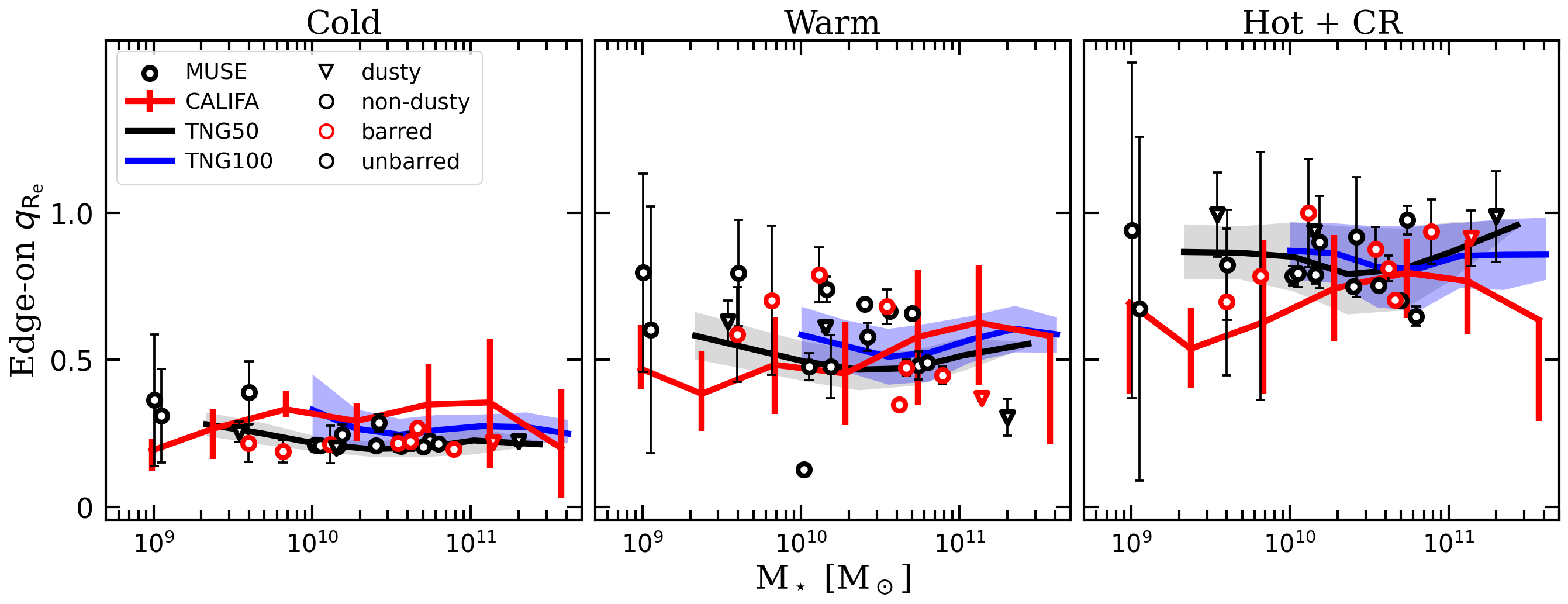}
        \caption{The edge-on flattening $q_{\rm Re}$ of cold, warm, and hot+CR components as a function of galaxy stellar mass $M_{\star}$, in comparison with the statistical results from CALIFA \citep{zhu2018stellar}, TNG50 \citep{zhang2025diverse} and TNG100 \citep{xu2019}. The symbols are same as \autoref{fig:decomposition}. Error bars on the MUSE points are the standard deviation of $q_{\rm Re}$ across 100 realisations of the kinematic Monte-Carlo procedure.}
        \label{fig:flat}
    \end{figure*}

    We further reconstruct the 3D spatial density distribution of the cold, warm, hot+CR components with their corresponding stellar orbits. In \autoref{fig:morp_comp}, we show the edge-on projected surface brightness of each component, illustrated by two typical galaxies in our sample. We further quantify the morphological shape of each component by measuring the flattening at the position of $R_{\mathrm{e}}$ ($q_{\rm R_e}$) from the edge-on surface brightness map. For each component we project the reconstructed three-dimensional density onto the $x$--$z$ plane at an inclination of $90\degr$ and take the isophote passing through $(x,z)=(R_{\mathrm{e}},0)$. The flattening $q_{R_{\mathrm{e}}}$ is the ratio of the semi-minor to the semi-major axis of that isophote, obtained from the second moments of the light it encloses, following the procedure used for CALIFA and TNG galaxies by \citet{zhu2018stellar} and \citet{zhang2025diverse}.

    In \autoref{fig:flat}, we compare the $q_{\rm R_e}$ we measured for these galaxies with the previous results from the CALIFA galaxies and the TNG50 and TNG100 galaxies \citep{zhu2018stellar, xu2019, zhang2025diverse}. In general, we have the average of $q_{\rm R_e}= 0.25, 0.6, 0.85$ for the cold, warm, and hot+CR components, respectively. There is no obvious variation in the flattening of each component as a function of galaxy stellar mass $M_{\star}$. In general, this is in agreement with the previous results from CALIFA galaxies. We have six galaxies with $M_{\star} \lesssim 10^{10}\,M_{\odot}$, and their hot components are as round as those in the more massive galaxies. Note that the warm components of the galaxies in our sample are also rounder than those of the CALIFA galaxies in the same mass range.

\section{Discussion}
\label{sec:discussion}

\subsection{Data and modelling limitations}

    The present sample provides a high-resolution anchor for interpreting larger but typically lower-spatial-resolution IFU surveys, such as SAMI and MaNGA. A key limitation of the present work is that the sample is not statistically representative of local galaxies. It is archive-selected, filtered by data quality, and further restricted by the suitability of each system for axisymmetric Schwarzschild modelling. As a result, the present sample is well suited for a pilot comparison with previous orbital-decomposition studies, but not for a new population-level calibration of the stellar-mass dependence of orbital fractions. This distinction is important because the CALIFA-based relations are derived from much larger samples, whereas the current MUSE sample primarily provides higher angular resolution for a smaller and more heterogeneous set of galaxies. Expanding the analysis to a statistically larger and more diverse sample, including barred, dusty, and more strongly asymmetric galaxies with controlled systematics, would clarify how robustly orbital fractions trace galaxy assembly across morphology and environment.

    A second limitation is that spectral resolution of MUSE remains limited. MUSE improves substantially on the spatial resolution and S/N of the large
    IFU surveys, but its spectral resolution restricts what can be recovered
    of $h_3$ and $h_4$ in the low-mass galaxies. Below the instrumental dispersion these moments are not well constrained: in most of the low-mass systems they return values close
    to zero, and only in the better-resolved central regions, such as the nuclear disc of NGC\,4941,
    are they individually informative. Maps of this kind are still worth including in
    the dynamical model, because values consistent with zero carrying large uncertainties prevent the
    fit from adopting extreme higher moments that the data do not support. The information they carry
    is small, however, and that is reflected in the results: the uncertainty propagates through the
    orbital weights and appears as the large error bars on the low-mass galaxies in
    \autoref{fig:decomposition} and \autoref{fig:flat}.

   A few galaxies in the sample are dust-rich and are marked in \autoref{tab:best_model_summary}. Rather than removing them we retain them, because, as set out in \autoref{subsec:sample}, their stellar kinematic maps are regular over the modelled field and their orbit solutions are therefore constrained in the same way as the rest of the sample. Although we have excluded the strong barred galaxies, we included the galaxies with a weak bar, which are also marked in \autoref{tab:best_model_summary}. We do not include the bar explicitly in the model, the pattern motion of the bar will not be perfectly fitted. However, we have tested in several previous papers that its general stellar orbit distribution, with the coarse cut of cold, warm, and hot orbits, can be well recovered, by a model not include the bar explicitly \citep{Zhu2018MNRAS.473.3000Z, zhu2018stellar, zhu2020disentangling}. Although recovering the detailed barred structures needs a model to construct the bar explicitly \citep{jin2025orbit}, which is beyond the scope of this paper.

\subsection{The warm component in low-mass galaxies}

    \begin{figure*}[ht!]
        \centering
        \includegraphics[width=\textwidth]{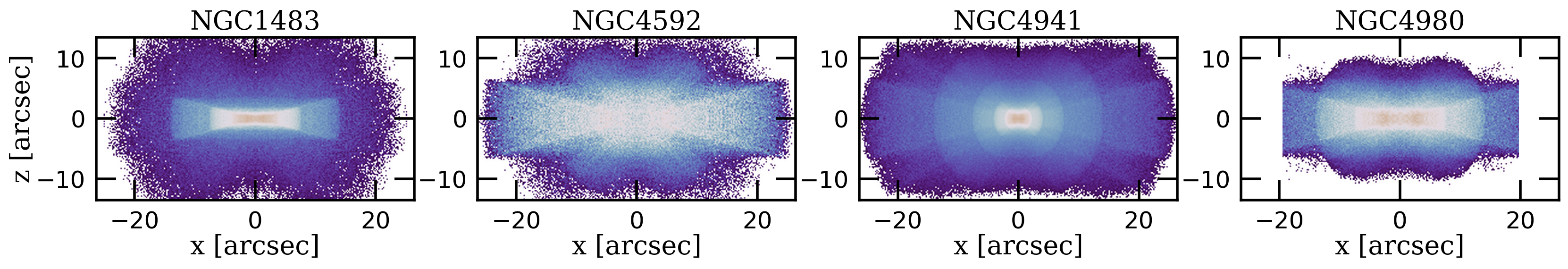}
        \caption{Edge-on projected surface-brightness distributions of the warm component in NGC\,1483, NGC\,4592, NGC\,4941 and NGC\,4980. The four systems shown here all have $f_{\rm warm}\geq0.59$.}
        \label{fig:warm_morph}
    \end{figure*}

 One notable aspect of the current sample is that a warm component is particularly prominent in several low-mass galaxies. Nevertheless, given the limited spectral resolution of the MUSE observations, the kinematic maps for these low-mass systems carry substantial uncertainties. This, in turn, leads to large uncertainties in the inferred orbital fractions (as shown in \autoref{fig:decomposition}). Additional observations with higher spectral resolution are still required to verify the orbital fractions in the low-mass galaxies. 
    
    In the CALIFA results and in IllustrisTNG, the warm component is not a pathological outlier at low mass but part of a broader distribution of orbital structures. Our MUSE sample is consistent with that picture while showing in more detail that the warm component can correspond to more than one apparent morphology. This immediately argues against treating the warm component as a one-to-one proxy for a single physical structure.
    
    In general, the warm component may include a mixture of vertically extended discs, centrally concentrated structures, bar-heated stellar populations, and intermediate-circularity stars that do not map cleanly onto classical bulge-versus-disc language. The present data are sufficient to show that low-mass galaxies need not be cold-disc dominated in terms of orbital fractions within $R_{\mathrm{e}}$, but they are not yet sufficient to determine a unique physical interpretation for the warm component on a galaxy-by-galaxy basis. A stronger follow-up analysis would quantify the intrinsic flattening and radial concentration of the warm orbits and compare them with independent indicators such as bar classification, photometric bulge prominence, and stellar population age. That would make the warm component scientifically interpretable rather than merely descriptive. Also, a useful next step is to test whether the warm-component excess in some low-mass galaxies correlates with stellar population age, metallicity, gas fraction, environment, or signatures of recent disturbance. Joint constraints from stellar populations, ionised-gas kinematics, and, where available, H\,{\sc i} or CO data would help distinguish between secular disc heating, minor mergers, feedback-driven turbulence, and inside-out disc settling as possible origins of the observed orbital mix. 
    
    The edge-on reconstructions in \autoref{fig:warm_morph} show that the warm orbits in the systems with the largest warm fractions are radially extended, forming thick discs rather than being confined to a compact central spheroid. They are, however, substantially thicker than the disc formed by the cold orbits. This is consistent with results from large samples of galaxies analysed by photometric decomposition, which show that the thick-to-thin disc mass ratio rises towards lower stellar mass \citep{yoachim2006structural,comeron2011thick}.

\subsection{Kinematics and evolutionary pathways}

    Our decomposition results map well onto the mass-dependent evolutionary framework recently proposed by \citet{zhang2025diverse}. In this scenario, low-mass galaxies largely maintain their intrinsic kinematics throughout their evolution, whereas the stellar orbits of high-mass galaxies are subjected to heating, either violently through major mergers or gradually via secular processes.

    This mass-dependent structural transition is reflected in our sample. In the low-mass regime ($M_{\star} \lesssim 10^{10.5}\,M_{\odot}$), most of our sampled galaxies exhibit minimal hot orbital fractions ($f_{\mathrm{hot}} \lesssim 0.2$). The luminosity fraction of the cold component initially increases with stellar mass, peaking near $M_{\star} \approx 10^{10.5}\,M_{\odot}$. Furthermore, our two lowest-mass systems fall along the low-mass extension of established cold component trends, confirming those broader statistical results.

    In the high-mass regime, although some galaxies still retain a low fraction of hot orbits, half of the galaxies in this mass bin exhibit $f_{\mathrm{hot}} > 0.2$, which is accompanied by a decrease in the cold component fraction as stellar mass increases. This divergence is consistent with the increasing dominance of orbital heating mechanisms at higher stellar masses.

    A direct comparison with cosmological simulations should move beyond intrinsic orbital fractions alone: mock observations that include the same spatial resolution, S/N selection, inclination distribution, and Schwarzschild modelling pipeline would make comparisons with simulations such as IllustrisTNG more stringent. Such forward modelling will be essential for turning dynamical decompositions into quantitative constraints on the formation histories of nearby disc galaxies.

\section{Conclusions}
\label{sec:conclusions}

We explored the MUSE archive data for nearby galaxies and picked 53 disc galaxies with relatively high S/N that allow for the extraction of stellar kinematic data. We performed pPXF analysis for the 53 galaxies, obtaining stellar kinematics including $v$, $\sigma$, $h_3$, and $h_4$. 

We further select 24 galaxies without strong dust obscuration or
pronounced non-axisymmetric structures, such as strong bars. We construct an orbit-superposition Schwarzschild model for these 24 galaxies and obtain their internal stellar orbit distributions. We dynamically decompose each galaxy into cold, warm, hot, and CR components, quantify their luminosity fractions as well as the edge-on flattening of each component. 

   The major results are as follows:
   \begin{itemize}
       \item The luminosity fractions of the cold, warm, hot, and CR components of these galaxies are broadly consistent with the mass-dependent trends previously reported from CALIFA, and with the TNG50 and TNG100 predictions.
       \item The average flattening of the cold, warm, and hot+CR components is $q_{\rm R_e}= 0.25, 0.6, 0.85$. We do not observe systematic variation in the flattening of the hot component as a function of stellar mass, as suggested by the CALIFA galaxies; the hot components in low-mass galaxies ($M_{\star} \lesssim 10^{10}\,M_{\odot}$) are as round as those in more massive galaxies.
       \item Three low-mass galaxies, including the two least massive in the sample, have $f_{\rm warm}\simeq0.6$, so that the dynamically warm component dominates the light within $R_{\mathrm{e}}$. The Monte-Carlo standard deviations on $f_{\rm warm}$ for the three highlighted galaxies are nevertheless 0.17, 0.21 and 0.084 (\autoref{fig:decomposition}); with only six low-mass galaxies, this result should be regarded as tentative.
   \end{itemize}

\begin{acknowledgements}
      The research presented here is supported by National Key R\&D Program of China No. 2022YFF0503403 (LZ), the NSFC key project No. 12633002 (LZ), and CAS Project for Young Scientists in Basic Research, Grant No. YSBR-062 (LZ), and the ESO, project number Ts~17/2--1 (JF). J.F-B acknowledges support from the PID2022-140869NB-I00 grant from the Spanish Ministry of Science and Innovation.
\end{acknowledgements}

%

\bibliographystyle{raa}
\bibliography{Ref} 

\onecolumn

\begin{appendix} 
\section{Basic Galaxy Properties}
\label{sec:appendix_a}

\autoref{tab:basic_info} presents the foundational properties of the 53 galaxies in the initial kinematic sample, including the adopted stellar mass $M_{\star}$ and its source, distance, and $\lambda_{R_{\mathrm{e}}}$. The data are sourced from the NASA/IPAC Extragalactic Database (NED) and \citet{erroz2019muse}. The effective radii $R_{\mathrm{e}}$ are taken directly from the literature, where they were measured by fitting elliptical isophotes to the galaxy images; we adopt the published values without refitting them here. The adopted stellar masses come from five published compilations, identified per galaxy in the last two columns of \autoref{tab:basic_info}: MAD \citep{erroz2019muse}; z0MGS \citep{leroy2019z0mgs}, from WISE $3.4\,\mu$m luminosities with a mass-to-light ratio set by WISE/GALEX colour; S$^4$G \citep{sheth2010s4g}, from Spitzer/IRAC $3.6\,\mu$m absolute magnitudes; AMUSING++ \citep{lopezcoba2020amusing}, from multi-stellar-population fitting of the integrated MUSE spectra with the \textsc{Pipe3D} pipeline and its GSD156 library, assuming a Salpeter IMF \citep{salpeter1955}; and the SDSS MPA-JHU catalogue (\texttt{galSpecExtra.lgm\_tot\_p50}), a Bayesian fit to SDSS photometry and spectroscopy.

\vspace{0.5cm}

\begin{longtable}{lccccc}
\caption{Basic Properties of the Galaxy Sample} \label{tab:basic_info} \\
\hline \hline
Galaxy & $R_{\mathrm{e}}$ & Distance & $\lambda_{R_{\mathrm{e}}}$ & $M_{\star}$ & Source \\
 & (kpc) & (Mpc) & & ($M_{\odot}$) & \\
\hline
\endfirsthead

\multicolumn{6}{c}%
{\tablename\ \thetable\ -- \textit{Continued from previous page}} \\
\hline \hline
Galaxy & $R_{\mathrm{e}}$ & Distance & $\lambda_{R_{\mathrm{e}}}$ & $M_{\star}$ & Source \\
 & (kpc) & (Mpc) & & ($M_{\odot}$) & \\
\hline
\endhead

\hline \multicolumn{6}{r}{\textit{Continued on next page}} \\
\endfoot

\hline
\endlastfoot

ESO339-G011   & 4.63       & 65.48       & 0.666       & $2.00 \times 10^{11}$            & AMUSING++ \\
ESO498-G5     & 2.39       & 31.90       & 0.517       & $1.04 \times 10^{10}$            & MAD \\
IC860         & 2.84       & 97.40       & 0.728       & $1.54 \times 10^{10}$            & MPA-JHU \\
IC2560        & 6.12       & 33.44       & 0.623       & $7.77 \times 10^{10}$            & MAD \\
IC5273        & 2.95       & 15.33       & 0.575       & $5.13 \times 10^{9}$             & z0MGS \\
NGC0289       & 3.70       & 19.16       & 0.591       & $4.60 \times 10^{10}$            & MAD \\
NGC0833       & 2.38       & 56.30       & 0.610       & $5.50 \times 10^{10}$            & z0MGS \\
NGC0835       & 0.96       & 34.00       & 0.547       & $5.50 \times 10^{10}$            & z0MGS \\
NGC1042       & 0.65       & 7.51        & 0.289       & $4.01 \times 10^{9}$             & S$^4$G \\
NGC1084       & 2.37       & 19.10       & 0.567       & $2.51 \times 10^{10}$            & MAD \\
NGC1097       & 9.87       & 17.16       & 0.503       & $1.74 \times 10^{11}$            & S$^4$G \\
NGC1300       & 7.24       & 14.79       & 0.627       & $3.77 \times 10^{10}$            & S$^4$G \\
NGC1309       & 2.77       & 28.13       & 0.435       & $1.12 \times 10^{10}$            & z0MGS \\
NGC1320       & 3.00       & 37.70       & 0.560       & $1.45 \times 10^{10}$            & z0MGS \\
NGC1326       & 2.74       & 14.95       & 0.473       & $4.17 \times 10^{10}$            & z0MGS \\
NGC1385       & 4.27       & 17.22       & 0.539       & $1.30 \times 10^{10}$            & S$^4$G \\
NGC1433       & 4.28       & 18.63       & 0.433       & $4.90 \times 10^{10}$            & z0MGS \\
NGC1483       & 2.22       & 21.23       & 0.448       & $6.52 \times 10^{9}$             & MAD \\
NGC1512       & 4.77       & 18.83       & 0.515       & $3.72 \times 10^{10}$            & z0MGS \\
NGC1566       & 3.19       & 17.69       & 0.502       & $4.68 \times 10^{10}$            & z0MGS \\
NGC1672       & 3.39       & 19.40       & 0.520       & $4.58 \times 10^{10}$            & S$^4$G \\
NGC1808       & 3.34       & 12.30       & 0.430       & $4.12 \times 10^{10}$            & S$^4$G \\
NGC2104       & 1.33       & 13.68       & 0.534       & $1.64 \times 10^{9}$             & MAD \\
NGC2835       & 3.44       & 10.07       & 0.606       & $6.28 \times 10^{9}$             & MAD \\
NGC3081       & 3.18       & 25.30       & 0.541       & $6.82 \times 10^{10}$            & MAD \\
NGC3256       & 6.13       & 37.40       & 0.367       & $1.38 \times 10^{11}$            & MAD \\
NGC3351       & 3.04       & 9.96        & 0.571       & $3.06 \times 10^{10}$            & S$^4$G \\
NGC3393       & 6.05       & 55.20       & 0.369       & $1.22 \times 10^{11}$            & MAD \\
NGC3513       & 2.29       & 12.33       & 0.256       & $2.34 \times 10^{9}$             & MAD \\
NGC3521       & 2.15       & 11.40       & 0.470       & $1.42 \times 10^{10}$            & MAD \\
NGC3627       & 3.62       & 11.32       & 0.620       & $6.44 \times 10^{10}$            & S$^4$G \\
NGC3783       & 4.81       & 39.60       & 0.397       & $4.08 \times 10^{10}$            & MAD \\
NGC4030       & 5.05       & 29.92       & 0.624       & $3.63 \times 10^{10}$            & z0MGS \\
NGC4303       & 4.09       & 16.99       & 0.519       & $4.68 \times 10^{10}$            & z0MGS \\
NGC4321       & 5.49       & 15.21       & 0.617       & $5.13 \times 10^{10}$            & z0MGS \\
NGC4418       & 1.69       & 22.97       & 0.646       & $3.47 \times 10^{9}$             & z0MGS \\
NGC4535       & 6.24       & 15.77       & 0.620       & $3.09 \times 10^{10}$            & z0MGS \\
NGC4592       & 0.85       & 11.59       & 0.550       & $1.12 \times 10^{9}$             & z0MGS \\
NGC4593       & 5.27       & 34.33       & 0.487       & $8.90 \times 10^{10}$            & MAD \\
NGC4603       & 8.68       & 35.37       & 0.621       & $4.68 \times 10^{10}$            & z0MGS \\
NGC4790       & 1.97       & 22.92       & 0.604       & $3.97 \times 10^{9}$             & MAD \\
NGC4900       & 3.52       & 21.53       & 0.308       & $1.74 \times 10^{10}$            & MAD \\
NGC4941       & 2.19       & 13.66       & 0.505       & $6.26 \times 10^{10}$            & MAD \\
NGC4980       & 1.86       & 17.00       & 0.498       & $1.01 \times 10^{9}$             & MAD \\
NGC5334       & 7.33       & 33.32       & 0.559       & $2.56 \times 10^{10}$            & S$^4$G \\
NGC5643       & 3.53       & 11.44       & 0.369       & $6.90 \times 10^{10}$            & MAD \\
NGC5806       & 2.25       & 24.63       & 0.596       & $5.02 \times 10^{10}$            & MAD \\
NGC6754       & 6.20       & 42.05       & 0.725       & $3.47 \times 10^{10}$            & z0MGS \\
NGC7162       & 2.87       & 34.95       & 0.715       & $2.63 \times 10^{10}$            & MAD \\
NGC7496       & 4.60       & 27.75       & 0.482       & $1.56 \times 10^{10}$            & MAD \\
NGC7552       & 2.35       & 11.14       & 0.269       & $1.55 \times 10^{10}$            & MAD \\
NGC7771       & 3.81       & 48.10       & 0.568       & $1.38 \times 10^{11}$            & z0MGS \\
PGC3853       & 1.89       & 11.53       & 0.397       & $6.08 \times 10^{9}$             & MAD \\

\end{longtable}

\subsection{Stellar mass from dynamical models}

    \begin{figure}
        \centering
        \includegraphics[width=\columnwidth]{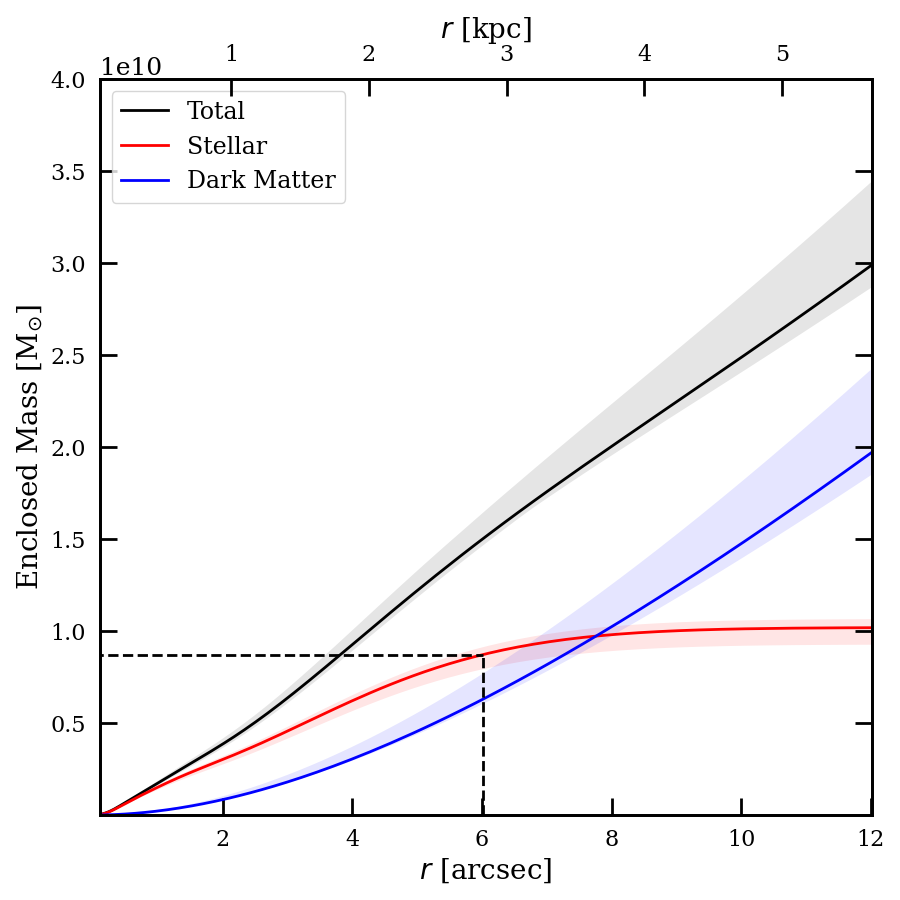}
        \caption{Spherically enclosed mass profiles as a function of radius derived from the best-fit dynamical model of IC\,860. The solid lines denote the total mass (black), the stellar component (red), and the DM halo (blue), with corresponding shaded regions representing the $1\sigma$ model uncertainties. The dashed lines mark $R_{\mathrm{e}}$ ($6.01\arcsec$) and the stellar mass enclosed within it ($10^{9.94}\,M_{\odot}$). Radii are shown in both angular (bottom) and physical (top) units.}
        \label{fig:mass_profile}
    \end{figure}

    To determine the total dynamical stellar mass for each galaxy in our sample, we utilise the enclosed mass profiles derived from our best-fit \textsc{dynamite} models. \autoref{fig:mass_profile} illustrates the spherically enclosed mass profiles for IC\,860 and decomposes the total mass into its stellar and DM components as a function of radius. The spatially resolved stellar mass is computed by taking the intrinsic luminosity density---which is recovered from the input MGE SB parameterisation---and multiplying it by the dynamically constrained $\Upsilon_{\mathrm{r}}$. The DM halo mass profile is then determined from $f_{200}$ and $C_{dh}$ of the best-fit model with an NFW DM halo assumption.

    Rather than claiming a formal total stellar mass from an extrapolation beyond the region securely constrained by the available imaging, we define an approximate dynamical stellar mass proxy based on the stellar mass enclosed within $R_{\mathrm{e}}$:
    \begin{equation}
        M_{\star-\mathrm{DYNAMITE}} = 2 \times M(<R_{\mathrm{e}})
    \end{equation}
    This definition assumes that $R_{\mathrm{e}}$ encloses roughly half of the total stellar light and is used here as a practical comparison quantity, not as an exact total-mass measurement. We adopt this choice because several archival images do not extend far enough to constrain the outer MGE components robustly; in those cases, integrating the MGE to a large radius would introduce a poorly controlled extrapolation. Where the imaging ends, the MGE is truncated rather than extrapolated: no outer Gaussian is added beyond the measured profile.

    \begin{figure}
        \centering
        \includegraphics[width=\columnwidth]{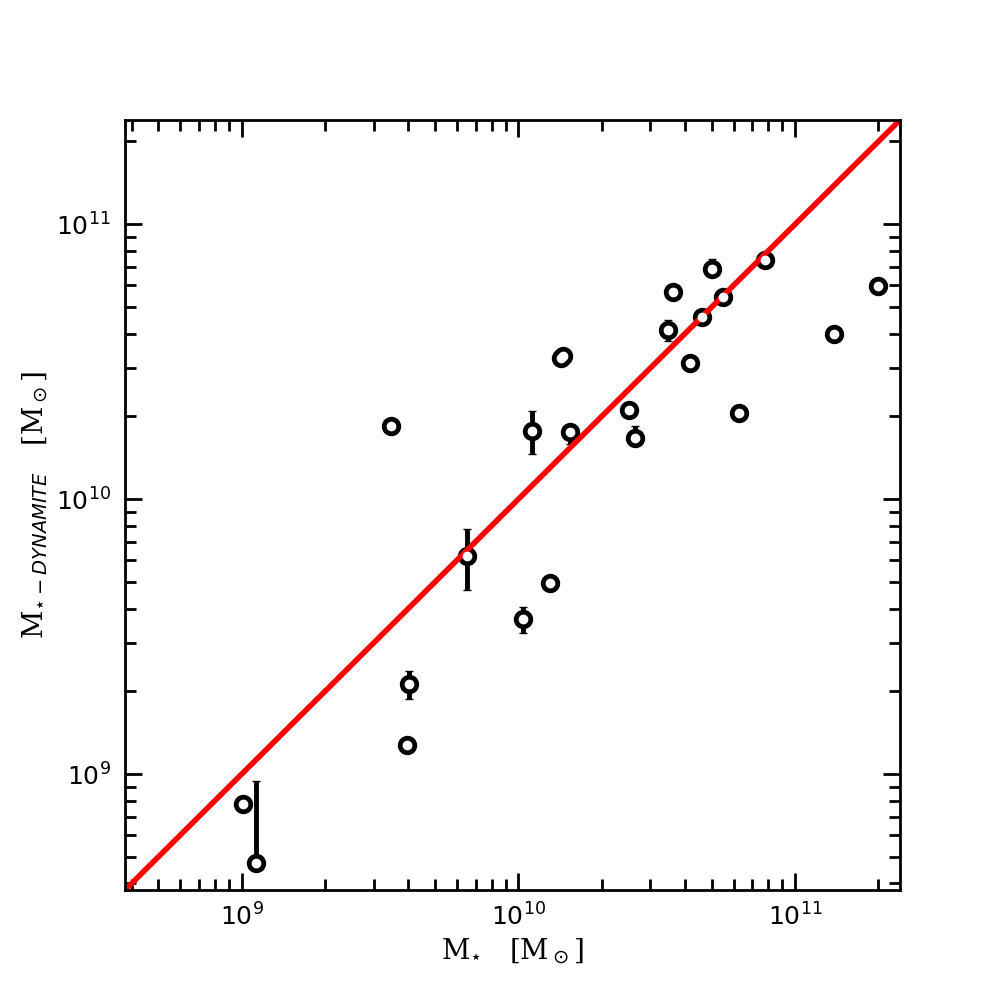}
        \caption{Comparison of the adopted stellar mass ($M_{\star}$) and the stellar mass from dynamical modelling with \textsc{dynamite} ($M_{\star - \text{DYNAMITE}}$). Each open black circle represents an individual galaxy. The solid red line indicates the one-to-one relation. $1\sigma$ model uncertainties are denoted with error bars.}
        \label{fig:mass_comparison}
    \end{figure}

    Finally, to validate our dynamical results, we compare our derived $M_{\star-\mathrm{DYNAMITE}}$ values against the independent stellar mass estimates $M_{\star}$ adopted in \autoref{tab:basic_info}. As shown in \autoref{fig:mass_comparison}, the two independent mass estimates are in good agreement across the sampled mass range. The two estimates agree to within a factor of $\sim2$ for 16 of the 24 galaxies, with a median $\log(M_{\star-\mathrm{DYNAMITE}}/M_{\star})$ of $-0.03$~dex and a scatter of 0.30~dex. Three galaxies are offset by more than 0.5~dex: $M_{\star-\mathrm{DYNAMITE}}$ exceeds $M_{\star}$ by 0.69~dex for NGC\,4418, and falls below $M_{\star}$ by 0.62~dex for NGC\,1042 and by 0.52~dex for ESO\,339-G011. The observed scatter remains within expected margins and can be attributed to standard methodological differences, such as variations in the assumed Initial Mass Function (IMF), or inherent degeneracies between the stellar mass and the DM halo profile in the inner regions of the dynamical models.

\section{Spectral resolution and the reliability of $h_3$ and $h_4$}
\label{subsec:lsf}

    Across the 5020--5450~\AA\ window the wavelength-dependent MUSE LSF of \citet{bacon2017muse} gives $\sigma_{\rm inst}=74$--$65$~km\,s$^{-1}$, or $69.7$~km\,s$^{-1}$ at 5200~\AA. This convolution limit is distinct from the $70$~km\,s$^{-1}$ sampling of the logarithmically rebinned spectra, although the two scales happen to be similar. Because a number of our bins have dispersions comparable to $\sigma_{\rm inst}$, we validated the extraction directly at the analysis configuration rather than relying on a nominal resolution criterion.

\begin{figure*}[ht!]
        \centering
        \includegraphics[width=\textwidth]{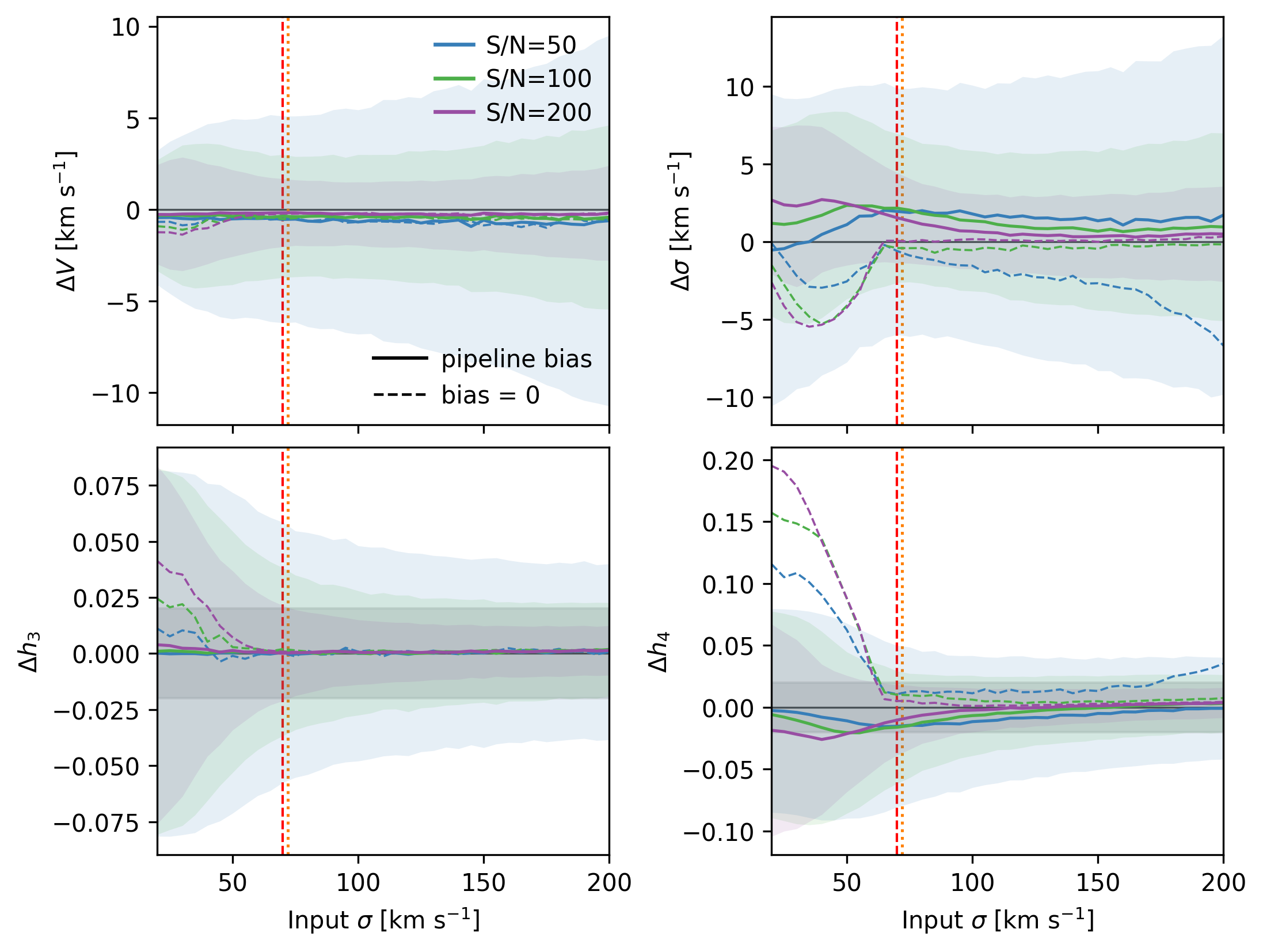}
        \caption{Monte-Carlo recovery for our pPXF set-up. Top row: the mean recovered-minus-input velocity and velocity dispersion. Bottom row: the same for the Gauss--Hermite coefficients $h_3$ and $h_4$. Lines show the mean after marginalising over the nine input $(h_3,h_4)$ combinations, and shaded regions show the $1\sigma$ scatter for the pipeline-default penalty; dashed lines show the mean recovery with the penalty set to zero. Colours denote S/N=50, 100 and 200 per pixel. Each grid cell contains 500 realisations. The dashed red and dotted orange vertical lines mark $\sigma_{\rm inst}=70$~km\,s$^{-1}$ and the velocity sampling of $70$~km\,s$^{-1}$, respectively; grey bands show the adopted $\pm 0.02$ bias tolerance for $h_3$ and $h_4$.}
        \label{fig:mc_test}
    \end{figure*}

    To test the reliability of $h_3$ and $h_4$ derived at this particular spectral resolution, we perform a Monte-Carlo test. We quantified recovery with 998,507 completed pPXF trials on a grid of 37 input dispersions from 20 to 200~km\,s$^{-1}$ in steps of 5~km\,s$^{-1}$, nine $(h_3,h_4)$ combinations, formed as the outer product of $h_3\in\{-0.10,0.00,+0.10\}$ with $h_4\in\{-0.10,0.00,+0.10\}$, three S/N values (50, 100 and 200 per pixel), 500 realisations and both the pipeline-default and zero-penalty settings. The mock spectra are sampled at the $70$~km\,s$^{-1}$ velocity scale of the data with templates sampled twice as finely, are fitted over 4870.7--5599.7~\AA\ with four Gauss--Hermite moments, additive and multiplicative polynomial degrees of $-1$ and $0$ and a free reddening, and the input velocity of each realisation is drawn uniformly over one velocity pixel so that the recovery is not evaluated at a preferred sub-pixel phase. The mocks use the MUSE-resolution optimal template of IC\,860 bin 13, integrate a Gauss--Hermite LOSVD on a ten-times-finer grid, randomise the sub-pixel velocity phase, adopt the wavelength dependence of that bin's real noise vector, and are refitted with the data configuration. The results are shown in \autoref{fig:mc_test}. The two moments that carry most of the constraint on the orbital weights are recovered essentially without bias over the entire range: at $\mathrm{S/N}=100$ the mean recovered-minus-input value is within $0.6$~km\,s$^{-1}$ for $V$ and $2.3$~km\,s$^{-1}$ for $\sigma$ at every grid point, with scatter below $5.0$ and $6.7$~km\,s$^{-1}$ respectively, including at dispersions well below $\sigma_{\rm inst}$. The higher moments behave differently. Averaged over the nine input $(h_3,h_4)$ combinations the mean bias at $\mathrm{S/N}=100$ stays within $0.002$ for $h_3$ and $0.021$ for $h_4$. That average is small, however, only because the input grid is symmetric about zero, so that the biases at $\pm0.10$ largely cancel. Taken cell by cell the penalty produces a systematic shrinkage of the higher moments towards zero: an input $h_3=0.1$ is recovered on average as $0.071$, and an input $h_4=0.1$ as $0.038$. This suppression is a designed property of the pPXF penalty rather than a failure of the fit, but it means that the higher moments of an individual spectrum are biased low, and that the recovery is not unbiased on a galaxy-by-galaxy basis. In addition, the scatter grows steeply as the dispersion falls below $\sigma_{\rm inst}$, so that below the instrumental dispersion $h_3$ and $h_4$ are not well constrained. What matters for this work is that this loss of information is reflected in the uncertainties: pPXF folds the scatter into the errors it reports on $h_3$ and $h_4$. Those uncertainties are the $\delta F$ term of $\chi^2_{\rm kin}$ in \autoref{eq:kinchi2}. Below the instrumental dispersion, therefore, $h_3$ and $h_4$ are best read as poorly constrained. The weak constraint is accounted for by the size of the uncertainties that accompany them. The uncertainties propagate into the orbital fractions through the Monte-Carlo procedure described in \autoref{subsec:decomp}, where they appear as the larger error bars carried by the affected galaxies in \autoref{fig:decomposition} and \autoref{fig:flat}. 

    As shown in \autoref{fig:mc_test}, above $\sigma_{\rm inst}$ the recovery of $h_3$ and $h_4$ improves with S/N, as expected, though below $\sigma_{\rm inst}$ it does not. The $\mathrm{S/N}=200$ curves are no better than, and for $h_4$ are worse than, those at $\mathrm{S/N}=50$. The reason is that in this regime the error budget of the higher moments is dominated by the penalty rather than by photon noise, and the weight of the penalty relative to the noise grows as the noise falls. Additional S/N therefore does not buy back the information lost to the instrumental convolution. We conclude that below the instrumental dispersion $h_3$ and $h_4$ are not well constrained, and that what protects the orbital decomposition is not the accuracy of these moments but the fact that their uncertainties are correspondingly large.
    
    Reporting $h_3$ and $h_4$ in this regime, with their uncertainties propagated, is nonetheless standard practice in integral-field spectroscopy. The SAURON survey operated at $\sigma_{\mathrm{inst}}=105$~km\,s$^{-1}$ \citep{bacon2001sauron}, half again larger than the $65$--$74$~km\,s$^{-1}$ of our fiducial window, and nonetheless published $h_3$ and $h_4$ maps for the bulges of 24 early-type spirals at a minimum $\mathrm{S/N}=60$ per resolution element \citep{falcon2006sauron}, for 18 late-type Sb--Sd spirals whose defining kinematic feature is a central dispersion drop \citep{ganda2006late}, and for Virgo and field dwarf ellipticals \citep{rys2013virgo}. \citet{rys2013virgo} state the trade-off explicitly: undersampling becomes important once the observed dispersion falls below roughly two spectral pixels, and it is compensated by correspondingly higher S/N. Closer to our set-up, \citet{gadotti2020kinematic} derive $V$, $\sigma$, $h_3$ and $h_4$ maps from MUSE data in the same blue spectral region, at $\mathrm{S/N}\simeq40$, for the nuclear discs of the TIMER sample.

\section{Photometry and MGE surface-brightness models}
\label{sec:appendix_mge}

    Every dynamical model in this work takes its stellar luminosity density from an MGE
    parameterisation of the observed SB distribution. This appendix documents the imaging used, the PSF
    adopted for each deconvolution, the fitting procedure, and the resulting fit quality for all 24 galaxies.

\subsection{Imaging data}

    We use archival imaging assembled galaxy by galaxy, prioritising the highest angular resolution
    available. For 23 of the 24 galaxies this is \textit{HST} imaging: WFPC2 for 14 galaxies, ACS/WFC for
    four, WFC3/UVIS for two, WFC3/IR for one, and NICMOS for two. The exception is ESO\,339-G011, for which no usable archival \textit{HST} frame covers the galaxy. For this galaxy we collapse the MUSE cube itself over $4750$--$7000$\,\AA\ to form a white-light image. The filters span F435W, F450W, F606W, F814W, F140W, F160W and
    F190N, and are listed per galaxy in \autoref{tab:mge_summary}.

    The relevant instrument characteristics are as follows. WFPC2 has three $800\times800$ Wide Field
    chips at $0\farcs1$\,pixel$^{-1}$ covering an L-shaped $\sim150\arcsec\times150\arcsec$ field, plus the
    Planetary Camera at $0\farcs0455$\,pixel$^{-1}$ over $36\arcsec\times36\arcsec$; archival drizzled
    products centred on the PC are sampled at $0\farcs05$\,pixel$^{-1}$. ACS/WFC covers
    $202\arcsec\times202\arcsec$ at $0\farcs05$\,pixel$^{-1}$, and WFC3/UVIS
    $162\arcsec\times162\arcsec$ at $0\farcs0396$\,pixel$^{-1}$. The NICMOS camera used here is NIC3 ($51\farcs2$ field,
    $0\farcs1$\,pixel$^{-1}$). The MUSE white-light image has $0\farcs2$ pixels over the
    $60\arcsec\times60\arcsec$ wide-field mosaic. The adopted sampling of each frame is given in \autoref{tab:mge_summary}.

\subsection{PSF determination}

    The MGE fit deconvolves the observed SB by an explicit PSF, so a PSF model is required for every
    frame. For all \textit{HST} images we generate a synthetic PSF with \textsc{Tiny Tim} v7.5
    \citep{krist2011tinytim}, computed for the specific camera, detector chip, filter and position of the
    galaxy on the detector, since the \textit{HST} PSF varies across the field. Where the archival product
    is drizzled, the model PSF is convolved with a square drop kernel matching the drizzle
    ($\mathrm{pixfrac}=1$) before use. The resulting PSF is itself parameterised as a circular MGE, which is
    the form \texttt{mge\_fit\_sectors} requires. The Gaussian-equivalent FWHM ranges from $0\farcs084$
    (WFPC2/PC) to $0\farcs33$ (NICMOS/NIC3), and is tabulated per galaxy. For ESO\,339-G011 the PSF is the atmospheric seeing of the MUSE observation, taken as a Gaussian with the delivered image quality recorded in the cube SGS header, $0\farcs58$.

\subsection{MGE fitting procedure}

    All 24 galaxies were fitted homogeneously with the same pipeline. The photometric centre,
    ellipticity and position angle are first measured with \texttt{find\_galaxy}. The SB is then sampled by \texttt{sectors\_photometry} along 19 sectors spanning one quadrant, exploiting the assumed axisymmetry to average azimuthally at fixed elliptical radius. Foreground stars, close companions, chip gaps and detector defects are excluded by a bad-pixel mask before sampling; two galaxies required explicit hand-built masks, ESO\,339-G011 (two bright foreground stars, one of which is $\sim70$ times brighter than the nucleus and would otherwise bias the centring) and NGC\,833 (compact sources superposed on the outer isophotes). Finally the sectors are fitted with \texttt{mge\_fit\_sectors}, allowing up to 12 Gaussians and imposing an outer power-law
    slope constraint. The adopted
    solutions use between four and ten Gaussians.

\subsection{Fit quality and radial coverage}

\begin{figure*}[p]
    \centering
    \includegraphics[width=0.93\textwidth]{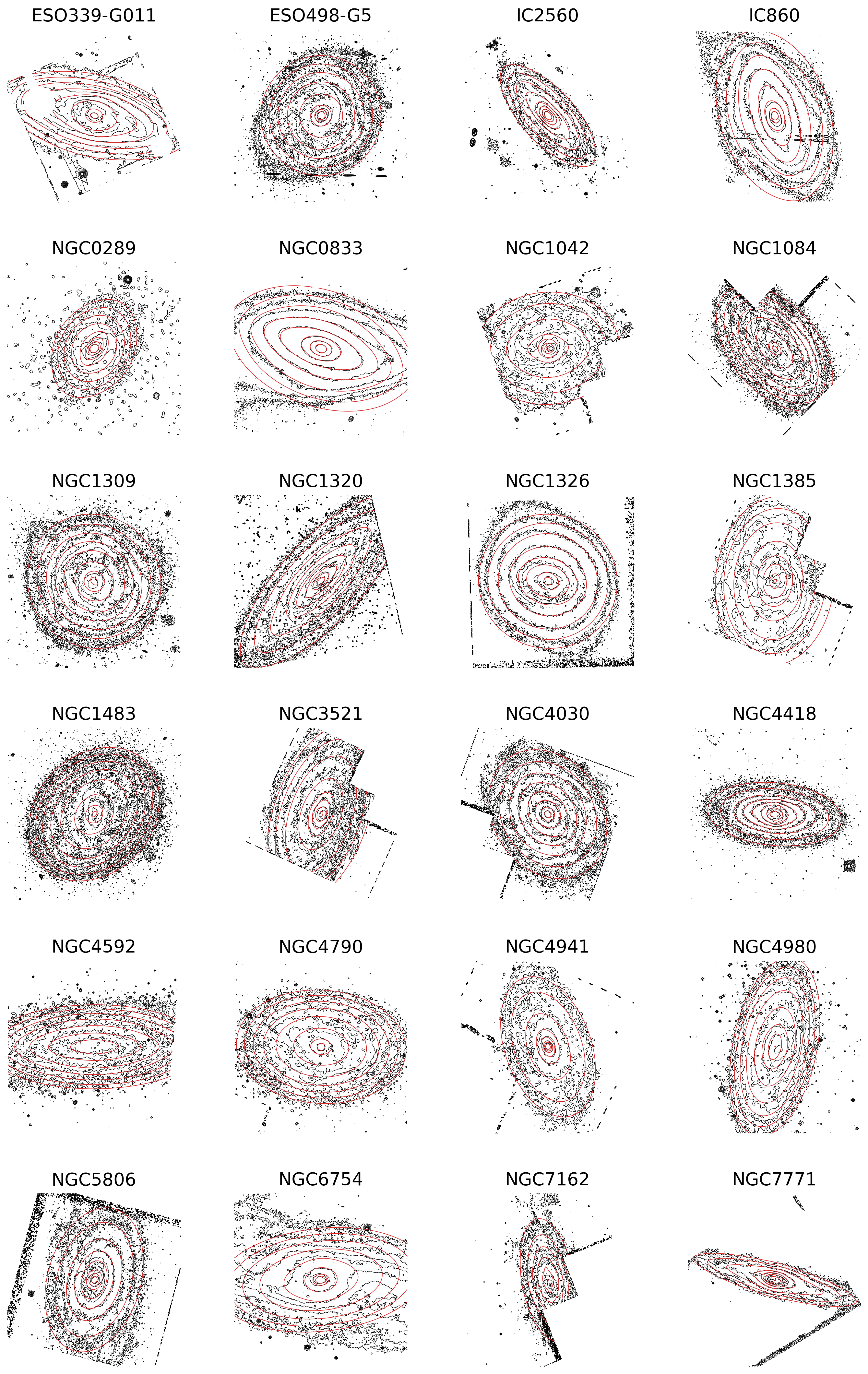}
    \caption{Observed and model isophotes for all 24 galaxies. Black contours show the observed
    photometry; red contours show the PSF-convolved MGE model evaluated at the same levels. Each panel is
    centred on the galaxy and oriented as observed; the irregular outlines reflect the archival detector
    footprints and the masked foreground sources described in the text.}
    \label{fig:mge_montage}
\end{figure*}

    \autoref{fig:mge_montage} compares the observed and model isophotes for all 24 galaxies, so that the quality of every stellar mass model used in this work can be judged directly. The agreement is generally close over the full fitted range. Quantitatively, the fractional Root Mean Square (RMS) residual between the model and the sector photometry ranges from 0.068 (NGC\,4790) to 0.263 (NGC\,6754).

    The photometry of seven galaxies is limited to less than $2\,R_{\mathrm{e}}$. Where the imaging runs out, the MGE is simply truncated at the outermost
    measured sector: no outer Gaussian is extrapolated beyond the data, and the luminosity density is therefore unconstrained outside that radius. This is the reason we do not quote a formally
    extrapolated total stellar mass anywhere in this work, and instead adopt
    $M_{\star-\mathrm{DYNAMITE}} = 2\,M(<R_{\mathrm{e}})$ as defined in \autoref{sec:appendix_a}. Because the orbital fractions reported in this paper are all measured within $R_{\mathrm{e}}$, they are insensitive to the unconstrained outer light.

\begin{table*}
    \centering
    \caption{Photometry and MGE fit summary for the 24 modelled galaxies. Columns give the instrument
    and filter of the frame used, its pixel sampling, the Gaussian-equivalent FWHM of the adopted PSF, the
    number of Gaussians in the final MGE, and the
    fractional RMS residual of the fit. Rows are ordered as in \autoref{tab:best_model_summary}.}
    \label{tab:mge_summary}
    \begin{tabular}{llccccc}
    \hline\hline
    Galaxy & Instrument & Filter & Sampling & PSF FWHM & $N_{\mathrm{G}}$ & RMS \\
     & & & (\arcsec/pix) & (\arcsec) & & \\
    \hline
ESO\,339-G011$^{d}$   & MUSE        & white light & 0.2000 & 0.580 & 6  & 0.189 \\
ESO\,498-G5     & ACS/WFC     & F814W       & 0.0500 & 0.119 & 5  & 0.166 \\
IC\,860   & WFC3/IR     & F140W       & 0.0500 & 0.146 & 6  & 0.092 \\
IC\,2560$^{b}$        & WFPC2       & F814W       & 0.1000 & 0.094 & 10 & 0.112 \\
NGC\,289$^{b}$  & WFPC2       & F606W       & 0.1000 & 0.195 & 5  & 0.164 \\
NGC\,833        & WFPC2       & F814W       & 0.1000 & 0.180 & 7  & 0.094 \\
NGC\,1042       & WFPC2       & F606W       & 0.1000 & 0.195 & 5  & 0.206 \\
NGC\,1084       & WFPC2       & F606W       & 0.1000 & 0.195 & 6  & 0.145 \\
NGC\,1309       & ACS/WFC     & F814W       & 0.0500 & 0.116 & 5  & 0.130 \\
NGC\,1320 & WFPC2       & F606W       & 0.0500 & 0.084 & 9  & 0.081 \\
NGC\,1326$^{b}$ & WFPC2       & F814W       & 0.0500 & 0.096 & 6  & 0.105 \\
NGC\,1385$^{b}$ & WFPC2       & F814W       & 0.1000 & 0.180 & 5  & 0.222 \\
NGC\,1483$^{b}$       & ACS/WFC     & F814W       & 0.0500 & 0.119 & 5  & 0.090 \\
NGC\,3521$^{d}$       & WFPC2       & F814W       & 0.1000 & 0.180 & 5  & 0.163 \\
NGC\,4030       & WFPC2       & F814W       & 0.1000 & 0.180 & 9  & 0.070 \\
NGC\,4418$^{d}$       & WFC3/UVIS   & F814W       & 0.0396 & 0.099 & 6  & 0.093 \\
NGC\,4592       & WFC3/UVIS   & F814W       & 0.0396 & 0.101 & 5  & 0.081 \\
NGC\,4790$^{b}$       & WFPC2       & F814W       & 0.1000 & 0.180 & 6  & 0.068 \\
NGC\,4941       & WFPC2       & F450W       & 0.1000 & 0.210 & 6  & 0.098 \\
NGC\,4980       & ACS/WFC     & F435W       & 0.0500 & 0.127 & 4  & 0.114 \\
NGC\,5806 & WFPC2       & F606W       & 0.0500 & 0.084 & 6  & 0.159 \\
NGC\,6754$^{b}$ & NICMOS/NIC3 & F160W       & 0.1000 & 0.294 & 4  & 0.263 \\
NGC\,7162       & WFPC2       & F606W       & 0.1000 & 0.195 & 5  & 0.112 \\
NGC\,7771$^{b,d}$       & NICMOS/NIC3 & F190N       & 0.1000 & 0.329 & 6  & 0.167 \\
    \hline
    \end{tabular}
    \vspace{8pt}
    \tablenotes{b}{\textwidth}{Weak barred galaxies.}
    \tablenotes{d}{\textwidth}{Dust-rich galaxies, as in \autoref{tab:best_model_summary}; see \autoref{subsec:sample}. }
\end{table*}

\section{Observed and modelled kinematic maps}
\label{sec:appendix_kinmaps}

\begin{figure*}[ht!]
        \centering
        \includegraphics[width=\hsize]{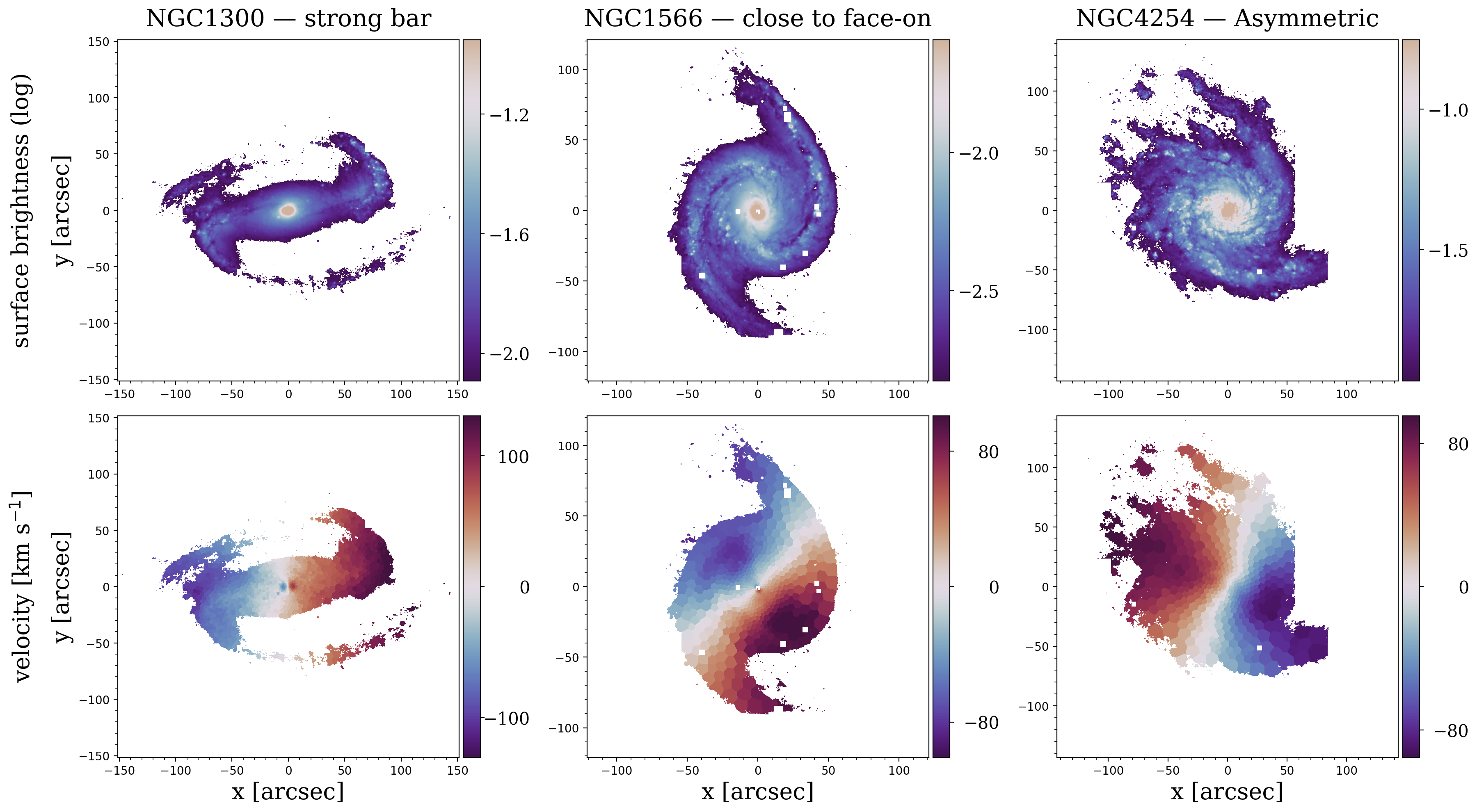}
        \caption{Examples of galaxies in the parent archive that were not modelled. Top: the
        reconstructed MUSE continuum flux, normalised to its peak. Bottom: the stellar line-of-sight
        velocity relative to the systemic value. NGC\,1300 is dominated by a strong bar, NGC\,1566 is close
        to face-on, and NGC\,4254 is markedly asymmetric in both its light distribution and its velocity
        field. All three violate the
        axisymmetry or data-quality requirements of our modelling, and no Schwarzschild models were built
        for them; the panels therefore show the observations rather than model fits.}
        \label{fig:rejected}
    \end{figure*}

Strongly barred and otherwise unsuitable galaxies are not modelled here. Examples from the MUSE archive are illustrated in \autoref{fig:rejected}: NGC\,1300, which hosts a strong bar; NGC\,1566, which is close to face-on and therefore provides little line-of-sight velocity information; and NGC\,4254, whose light distribution and velocity field are both markedly asymmetric.

    The Schwarzschild models are constrained by the full set of Voronoi-binned stellar kinematics,
    $V$, $\sigma$, $h_3$ and $h_4$, rather than by any radially averaged profile. To allow the quality of
    that fit to be judged directly, this appendix compares the observed maps with the corresponding
    best-fitting model predictions for every galaxy in the modelled sample, plotted on identical colour
    scales so that any systematic deviation is immediately visible. The galaxies appear in the same order
    as in \autoref{tab:best_model_summary}; IC\,860 is not repeated here, since its maps are
    already shown in \autoref{fig:model_example}.

    The model $h_3$ and $h_4$ maps are
    smoother than the data, as expected, since the observed maps retain the bin-to-bin scatter of the pPXF measurements while the model predictions are noise-free.

\begin{figure*}[p]
    \centering
    \includegraphics[width=\hsize]{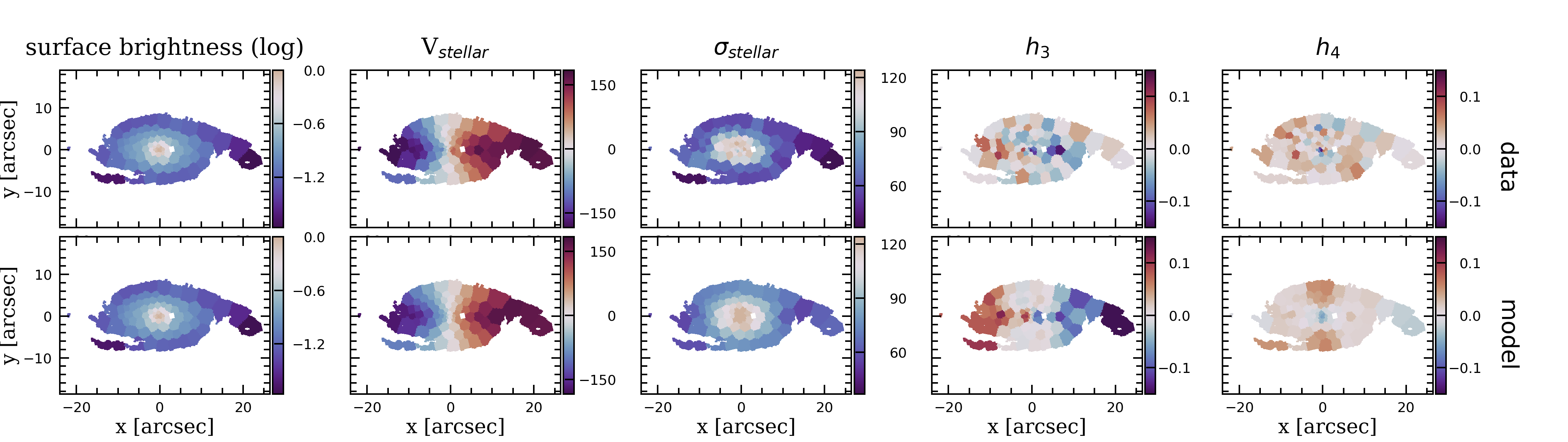}
    \caption{Observed and modelled stellar kinematics for ESO\,339-G011. Columns show,
    from left to right, the logarithmic SB, the mean line-of-sight velocity $V$, the velocity
    dispersion $\sigma$, and the Gauss--Hermite coefficients $h_3$ and $h_4$. The top row shows the
    Voronoi-binned MUSE measurements and the bottom row the corresponding prediction of the
    best-fitting Schwarzschild model, on identical colour scales.}
    \label{fig:kinmap_ESO339-G011}
\end{figure*}

\begin{figure*}[p]
    \centering
    \includegraphics[width=\hsize]{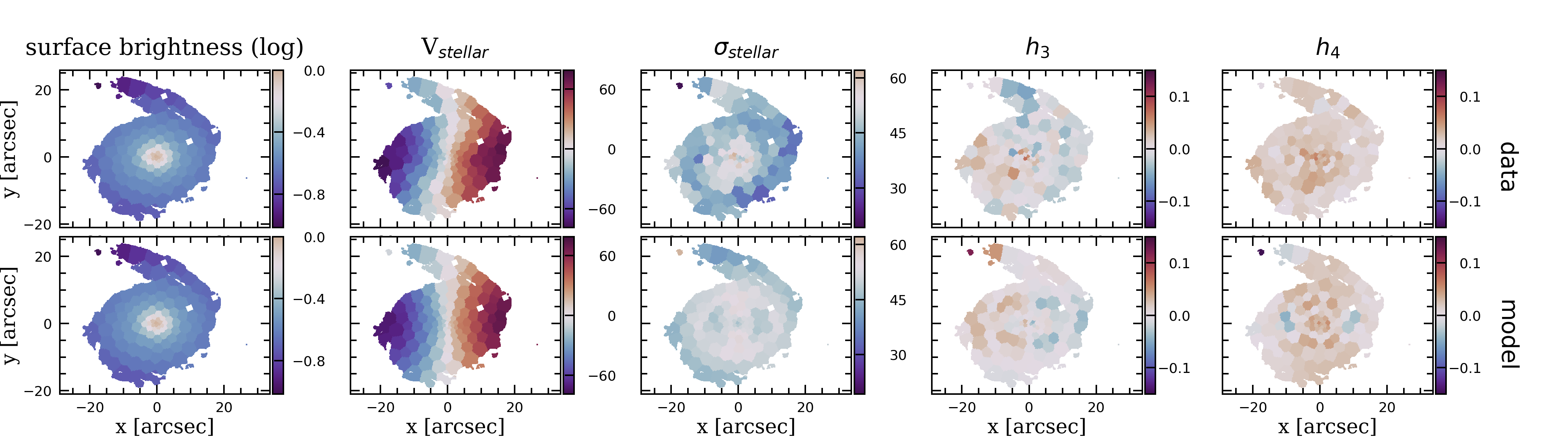}
    \caption{Same as \autoref{fig:kinmap_ESO339-G011}, but for ESO\,498-G5.}
    \label{fig:kinmap_ESO498-G5}
\end{figure*}

\begin{figure*}[p]
    \centering
    \includegraphics[width=\hsize]{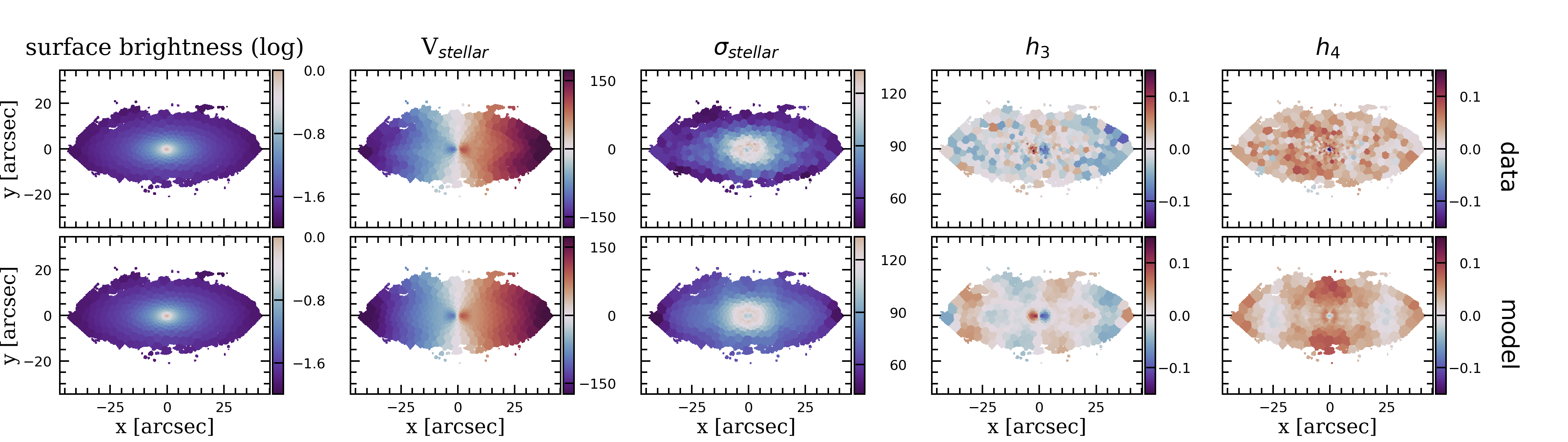}
    \caption{Same as \autoref{fig:kinmap_ESO339-G011}, but for IC\,2560.}
    \label{fig:kinmap_IC2560}
\end{figure*}

\begin{figure*}[p]
    \centering
    \includegraphics[width=\hsize]{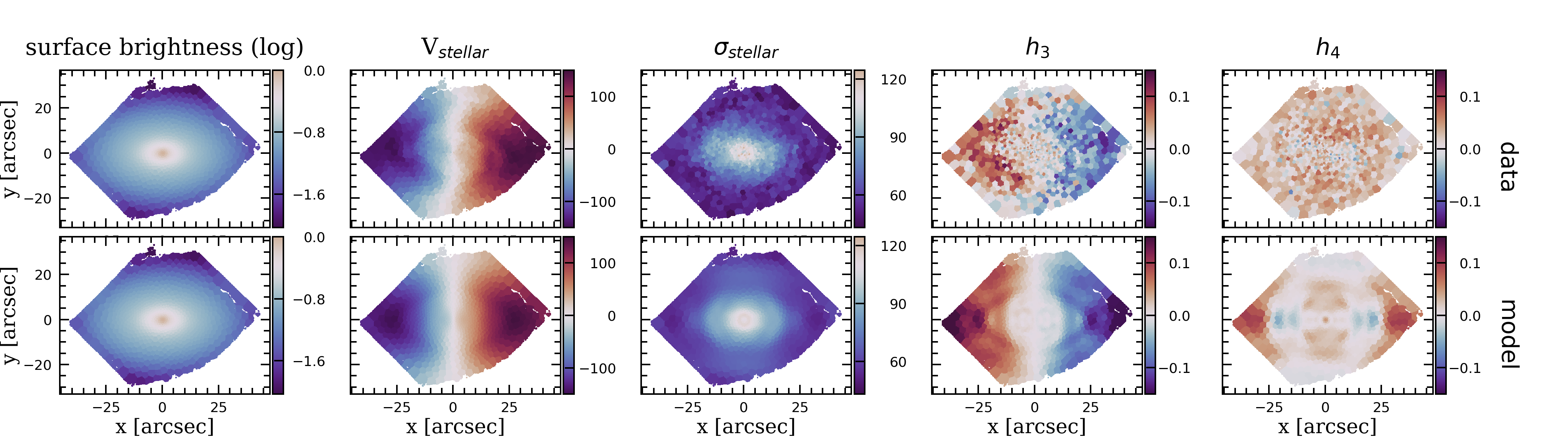}
    \caption{Same as \autoref{fig:kinmap_ESO339-G011}, but for NGC\,289.}
    \label{fig:kinmap_NGC0289}
\end{figure*}

\begin{figure*}[p]
    \centering
    \includegraphics[width=\hsize]{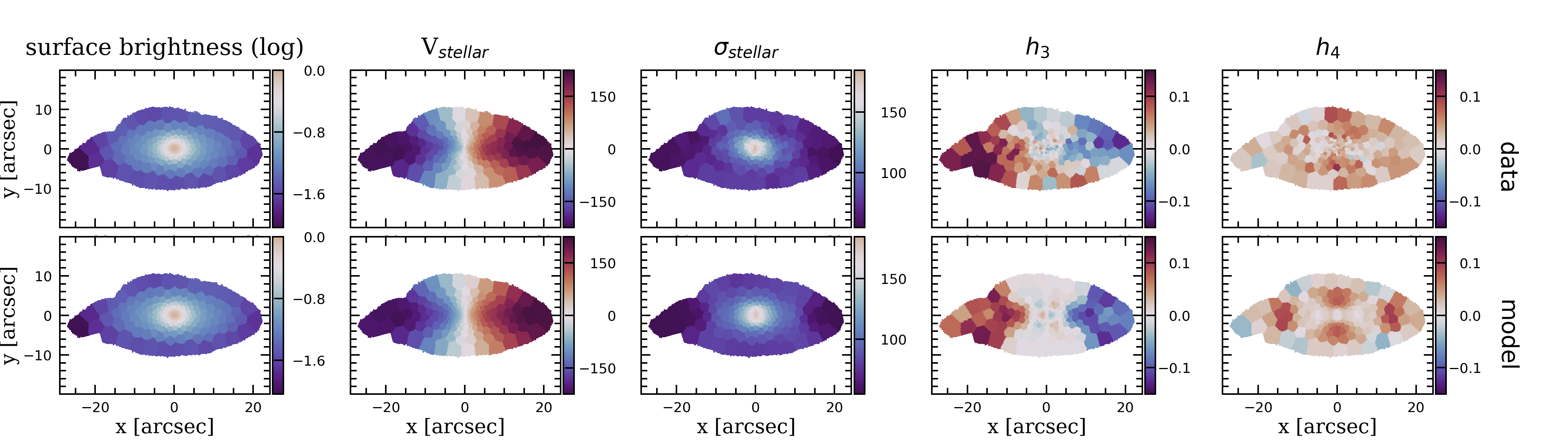}
    \caption{Same as \autoref{fig:kinmap_ESO339-G011}, but for NGC\,833.}
    \label{fig:kinmap_NGC0833}
\end{figure*}

\begin{figure*}[p]
    \centering
    \includegraphics[width=\hsize]{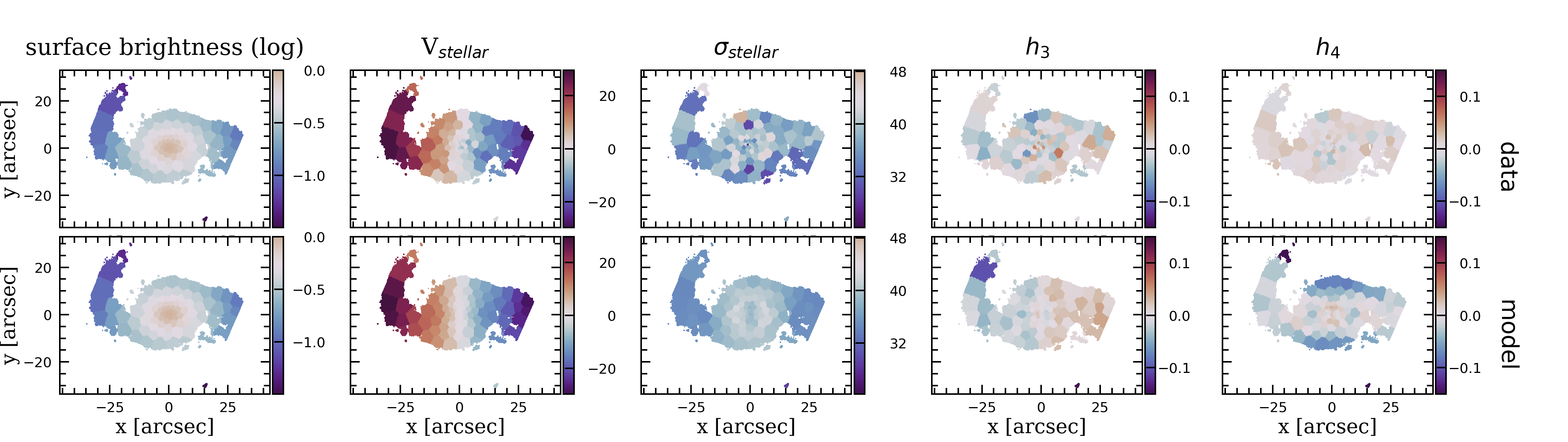}
    \caption{Same as \autoref{fig:kinmap_ESO339-G011}, but for NGC\,1042.}
    \label{fig:kinmap_NGC1042}
\end{figure*}

\begin{figure*}[p]
    \centering
    \includegraphics[width=\hsize]{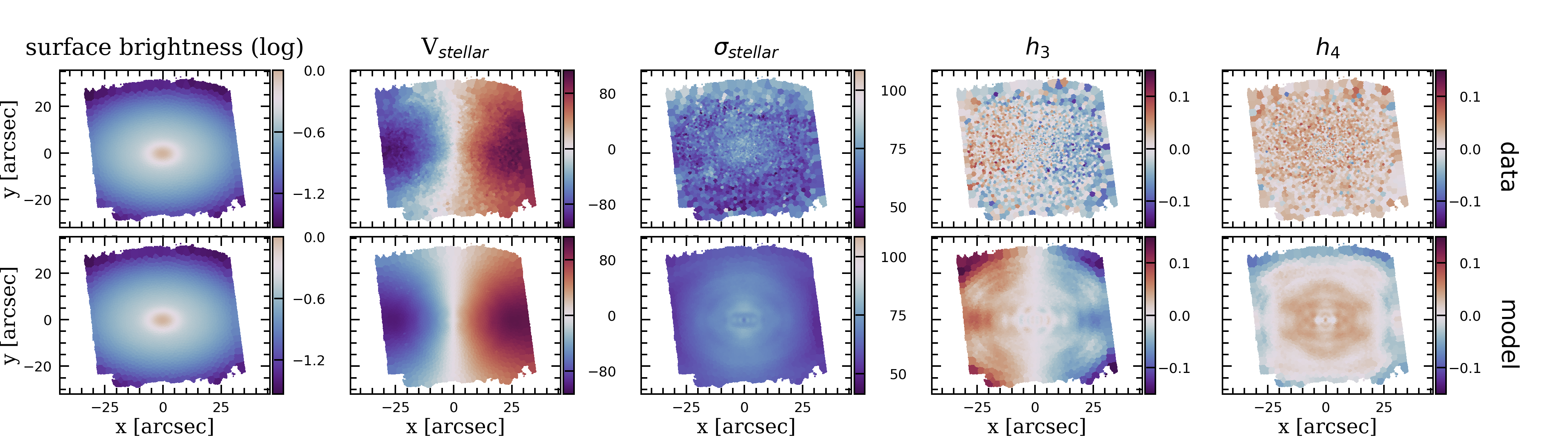}
    \caption{Same as \autoref{fig:kinmap_ESO339-G011}, but for NGC\,1084.}
    \label{fig:kinmap_NGC1084}
\end{figure*}

\begin{figure*}[p]
    \centering
    \includegraphics[width=\hsize]{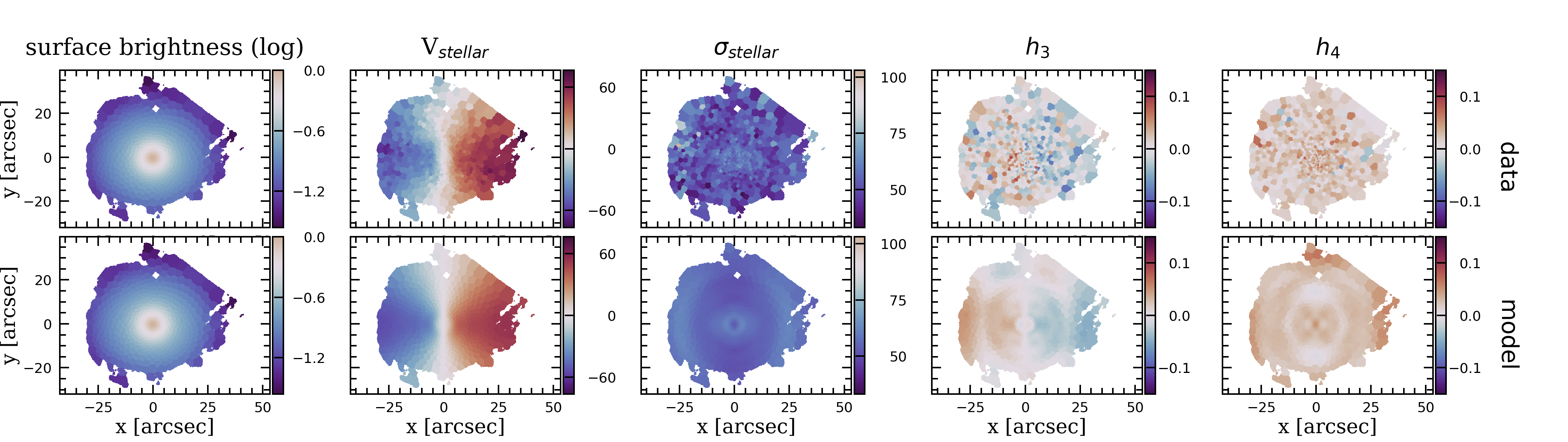}
    \caption{Same as \autoref{fig:kinmap_ESO339-G011}, but for NGC\,1309.}
    \label{fig:kinmap_NGC1309}
\end{figure*}

\begin{figure*}[p]
    \centering
    \includegraphics[width=\hsize]{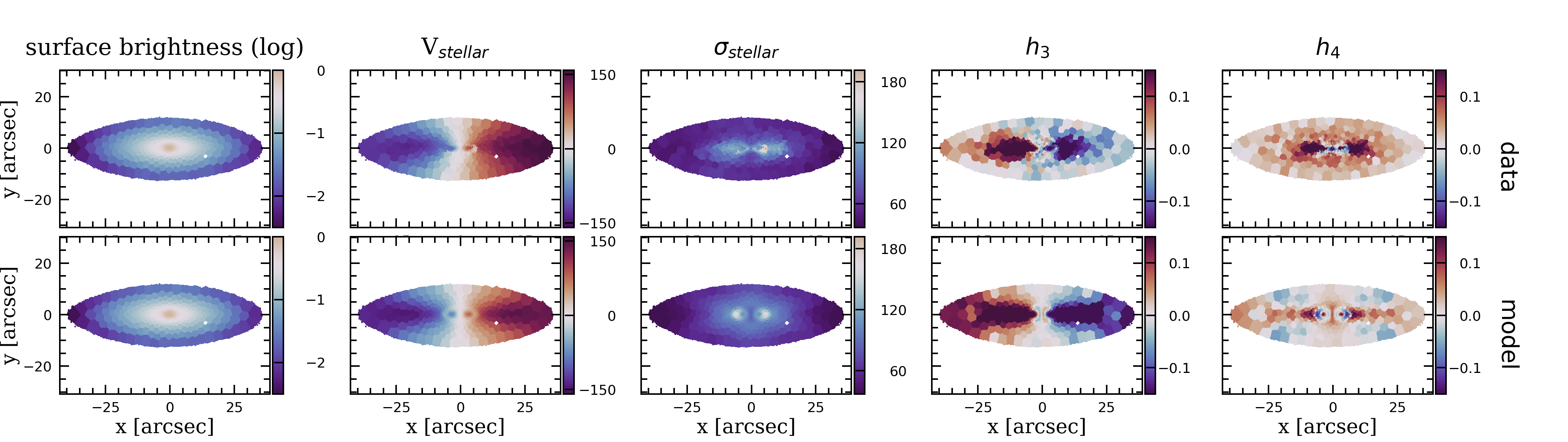}
    \caption{Same as \autoref{fig:kinmap_ESO339-G011}, but for NGC\,1320.}
    \label{fig:kinmap_NGC1320}
\end{figure*}

\begin{figure*}[p]
    \centering
    \includegraphics[width=\hsize]{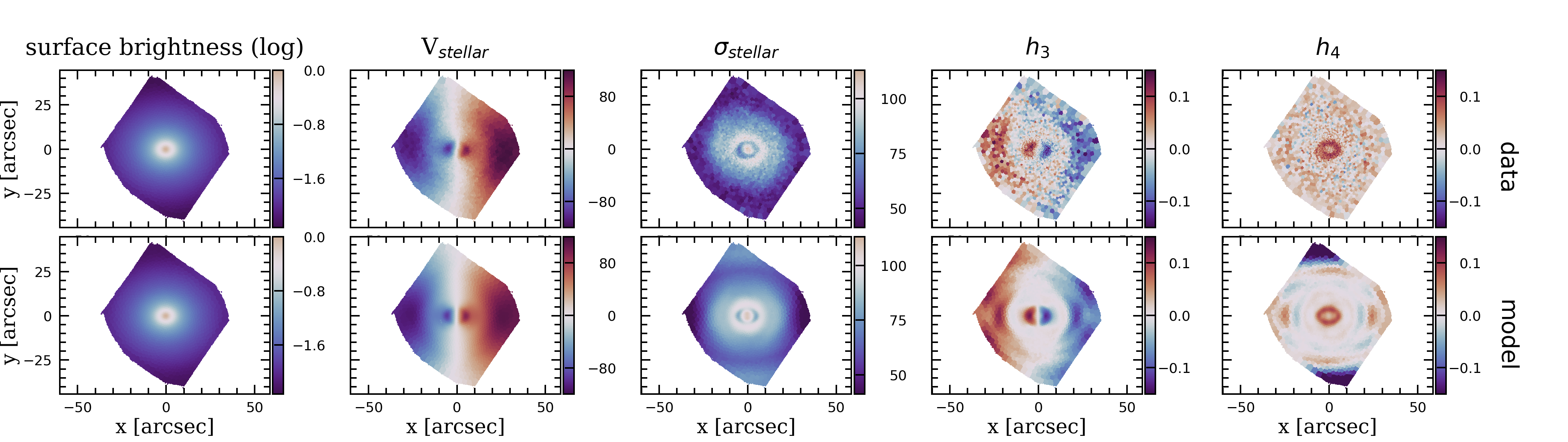}
    \caption{Same as \autoref{fig:kinmap_ESO339-G011}, but for NGC\,1326.}
    \label{fig:kinmap_NGC1326}
\end{figure*}

\begin{figure*}[p]
    \centering
    \includegraphics[width=\hsize]{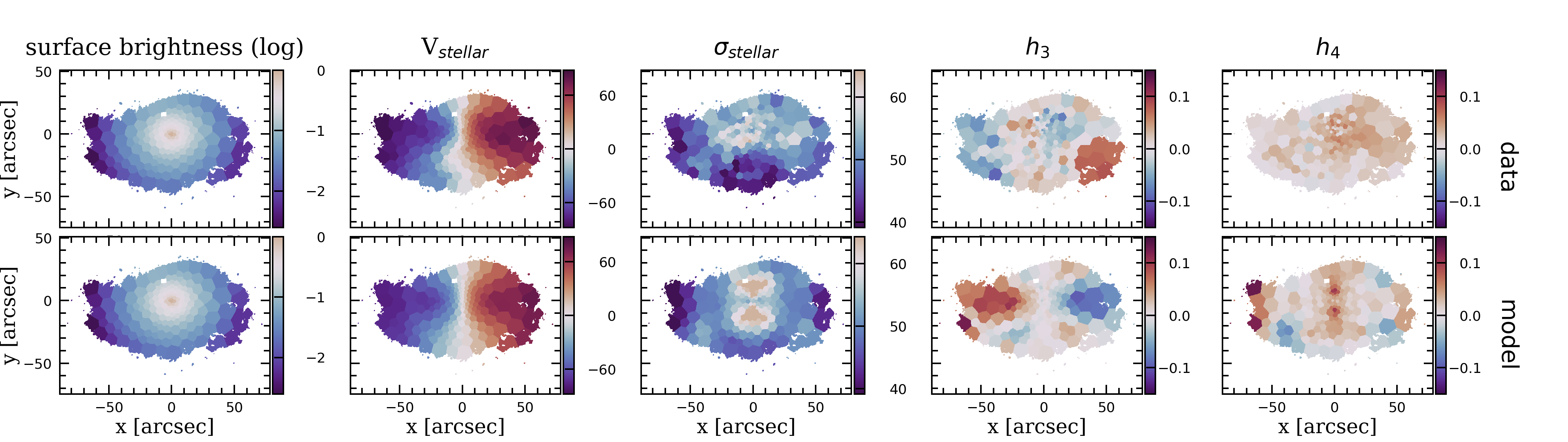}
    \caption{Same as \autoref{fig:kinmap_ESO339-G011}, but for NGC\,1385.}
    \label{fig:kinmap_NGC1385}
\end{figure*}

\begin{figure*}[p]
    \centering
    \includegraphics[width=\hsize]{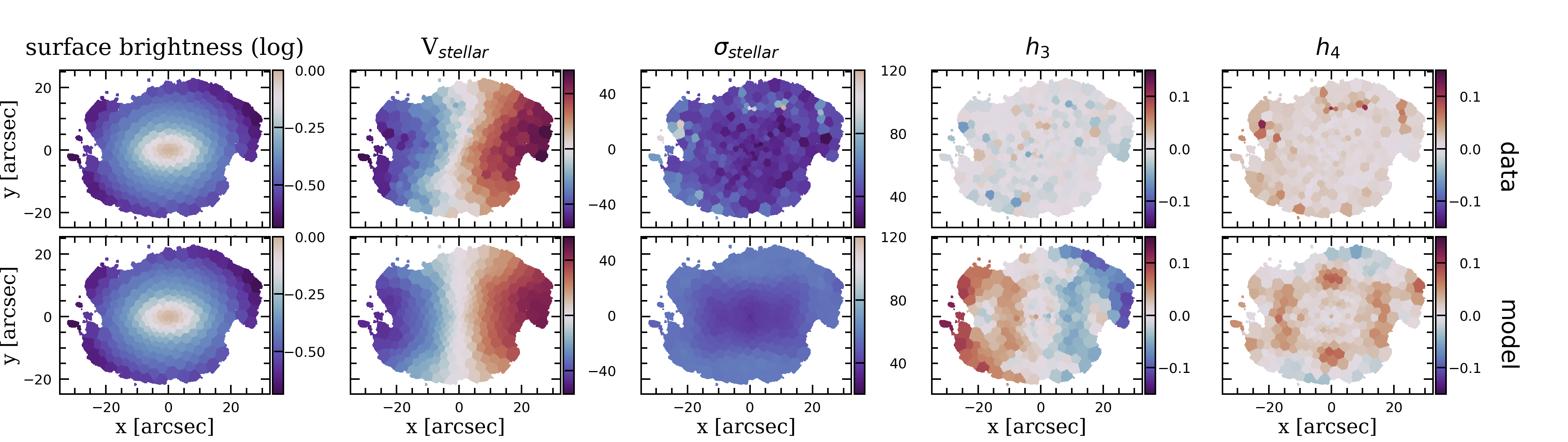}
    \caption{Same as \autoref{fig:kinmap_ESO339-G011}, but for NGC\,1483.}
    \label{fig:kinmap_NGC1483}
\end{figure*}

\begin{figure*}[p]
    \centering
    \includegraphics[width=\hsize]{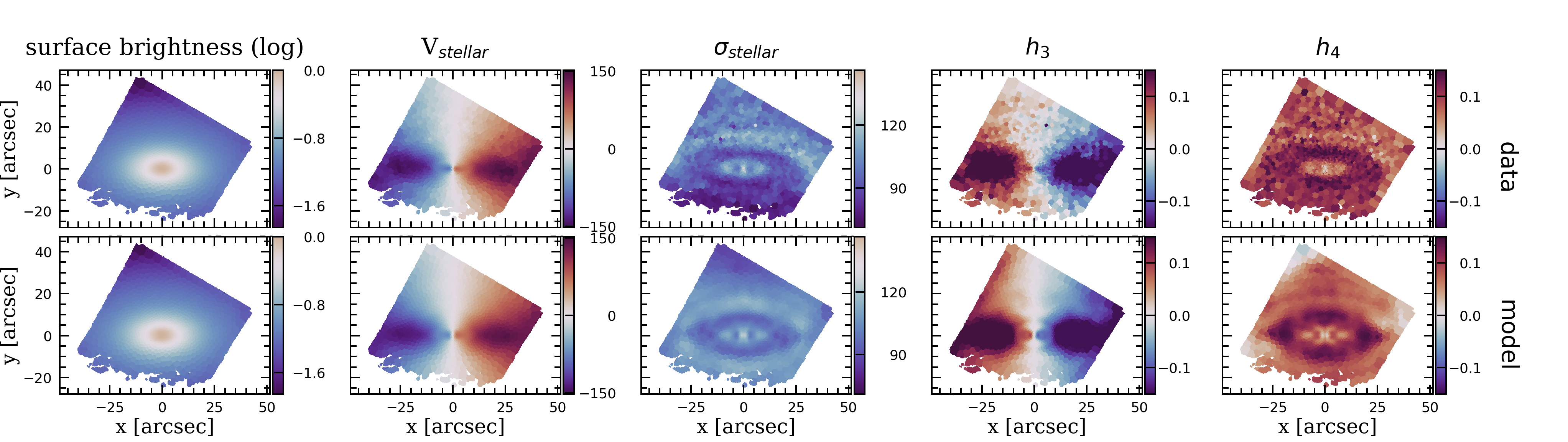}
    \caption{Same as \autoref{fig:kinmap_ESO339-G011}, but for NGC\,3521.}
    \label{fig:kinmap_NGC3521}
\end{figure*}

\begin{figure*}[p]
    \centering
    \includegraphics[width=\hsize]{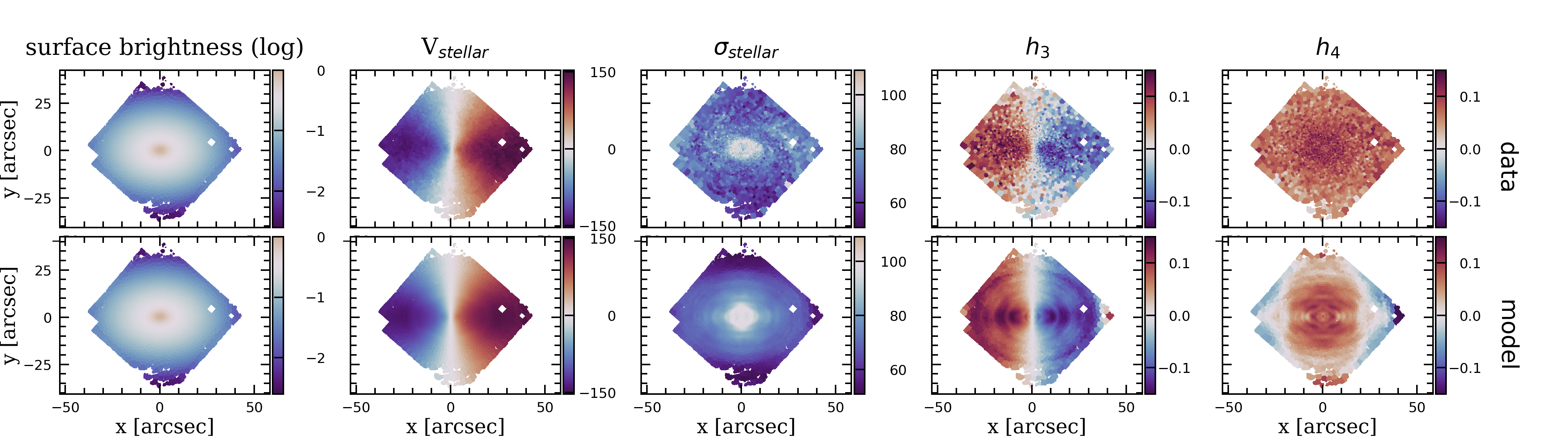}
    \caption{Same as \autoref{fig:kinmap_ESO339-G011}, but for NGC\,4030.}
    \label{fig:kinmap_NGC4030}
\end{figure*}

\begin{figure*}[p]
    \centering
    \includegraphics[width=\hsize]{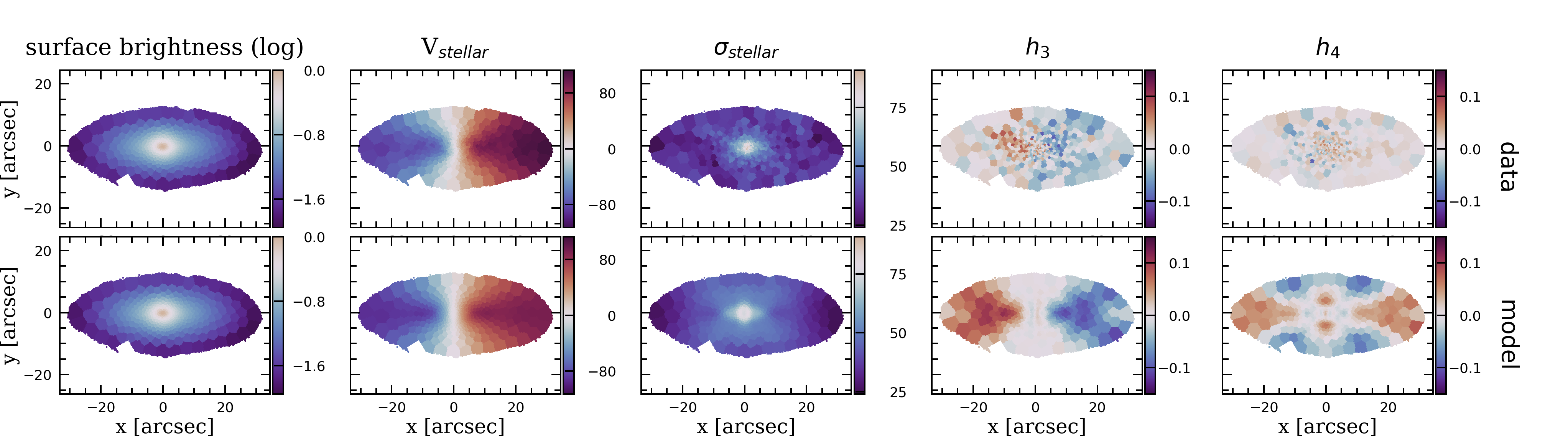}
    \caption{Same as \autoref{fig:kinmap_ESO339-G011}, but for NGC\,4418.}
    \label{fig:kinmap_NGC4418}
\end{figure*}

\begin{figure*}[p]
    \centering
    \includegraphics[width=\hsize]{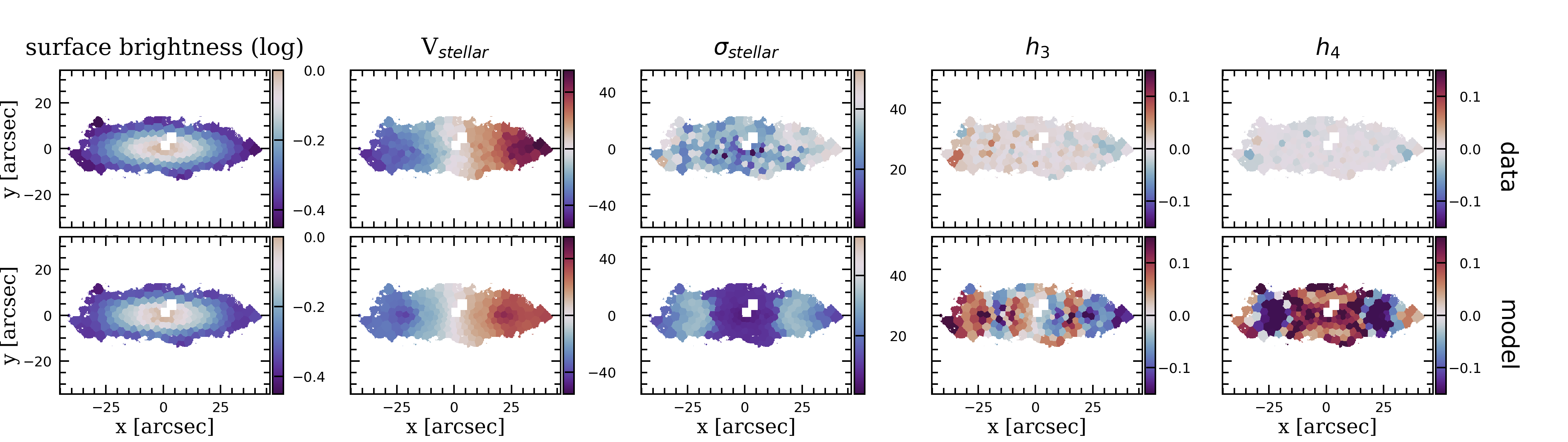}
    \caption{Same as \autoref{fig:kinmap_ESO339-G011}, but for NGC\,4592.}
    \label{fig:kinmap_NGC4592}
\end{figure*}

\begin{figure*}[p]
    \centering
    \includegraphics[width=\hsize]{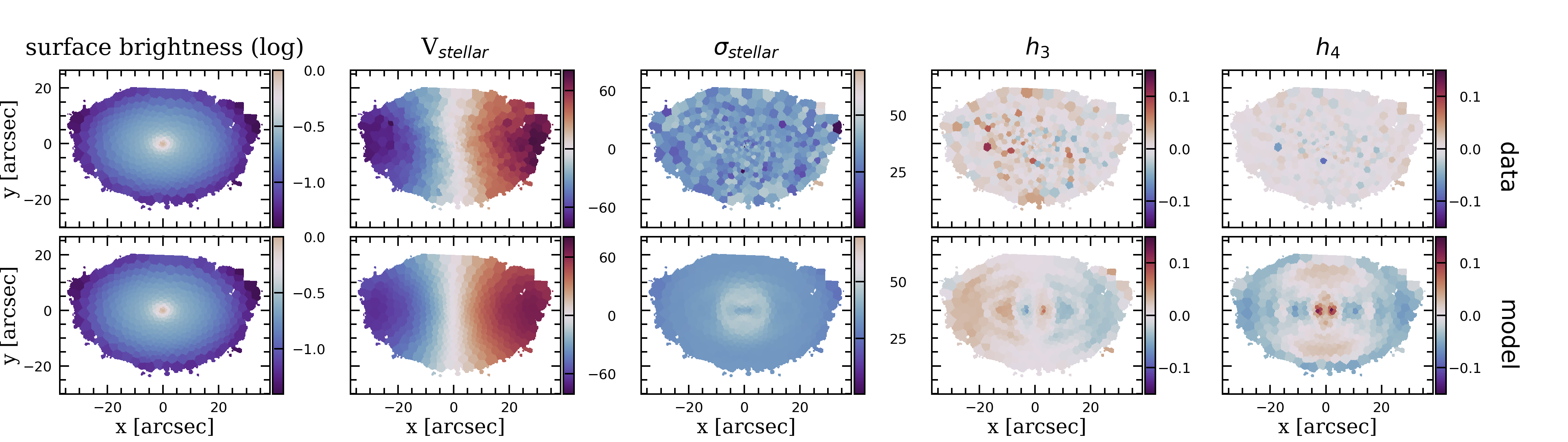}
    \caption{Same as \autoref{fig:kinmap_ESO339-G011}, but for NGC\,4790.}
    \label{fig:kinmap_NGC4790}
\end{figure*}

\begin{figure*}[p]
    \centering
    \includegraphics[width=\hsize]{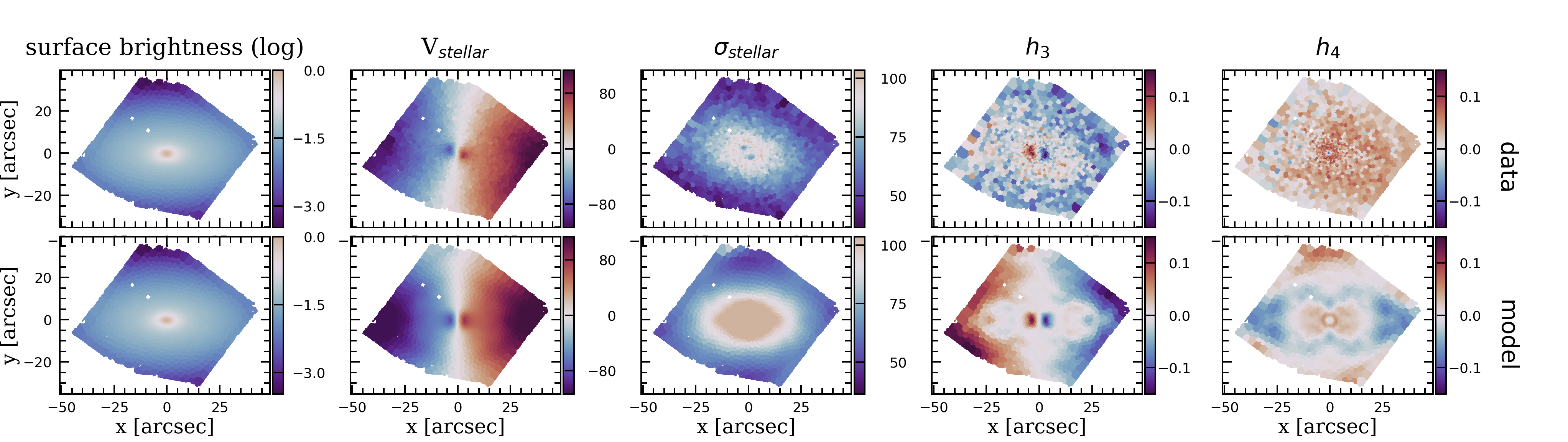}
    \caption{Same as \autoref{fig:kinmap_ESO339-G011}, but for NGC\,4941.}
    \label{fig:kinmap_NGC4941}
\end{figure*}

\begin{figure*}[p]
    \centering
    \includegraphics[width=\hsize]{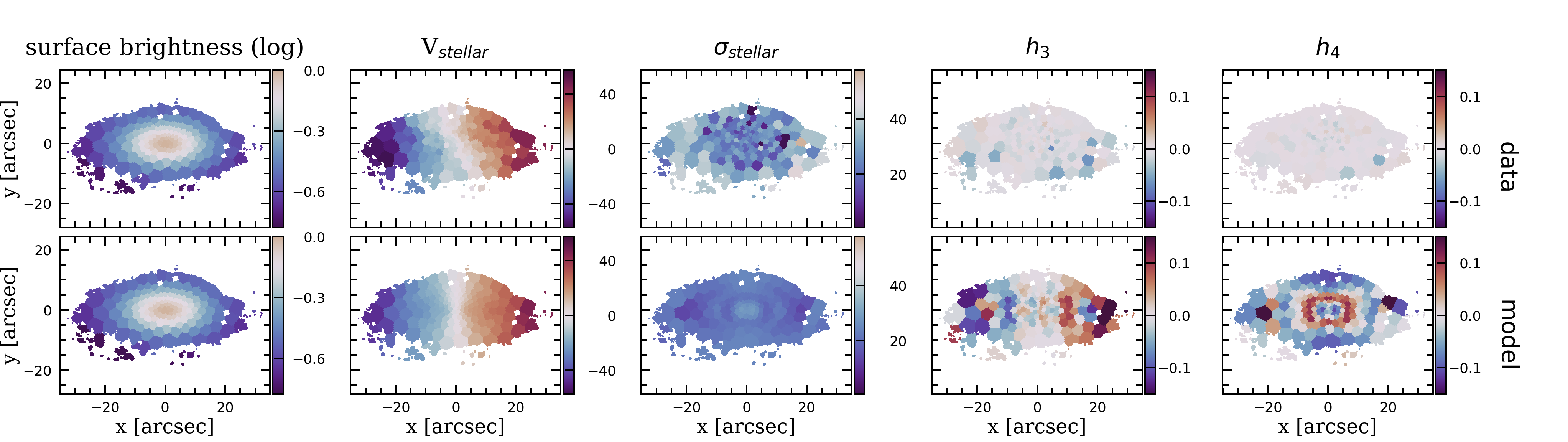}
    \caption{Same as \autoref{fig:kinmap_ESO339-G011}, but for NGC\,4980.}
    \label{fig:kinmap_NGC4980}
\end{figure*}

\begin{figure*}[p]
    \centering
    \includegraphics[width=\hsize]{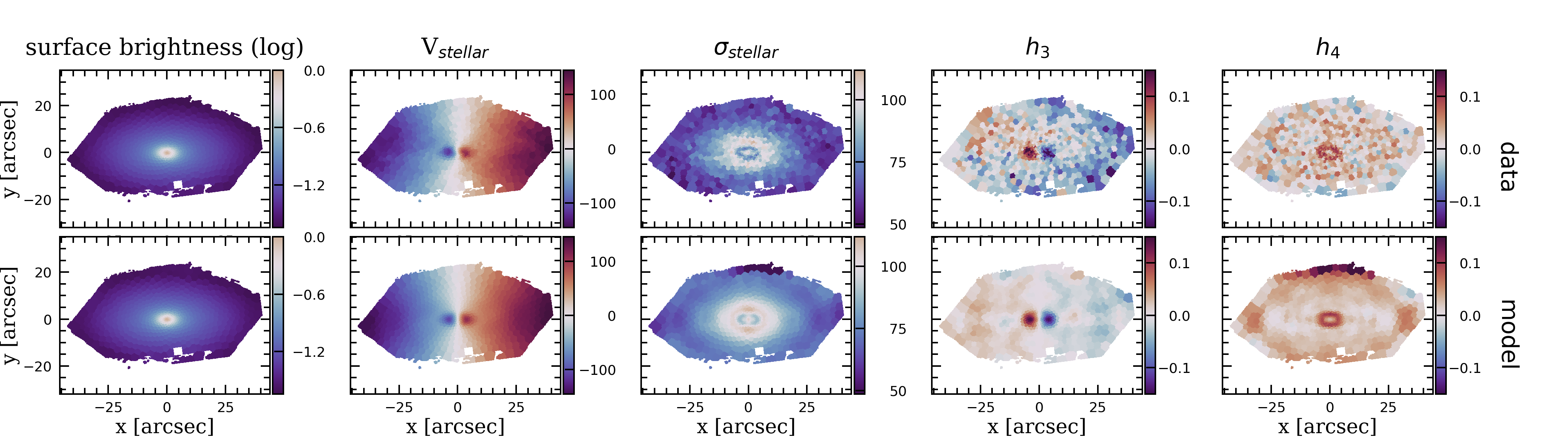}
    \caption{Same as \autoref{fig:kinmap_ESO339-G011}, but for NGC\,5806.}
    \label{fig:kinmap_NGC5806}
\end{figure*}

\begin{figure*}[p]
    \centering
    \includegraphics[width=\hsize]{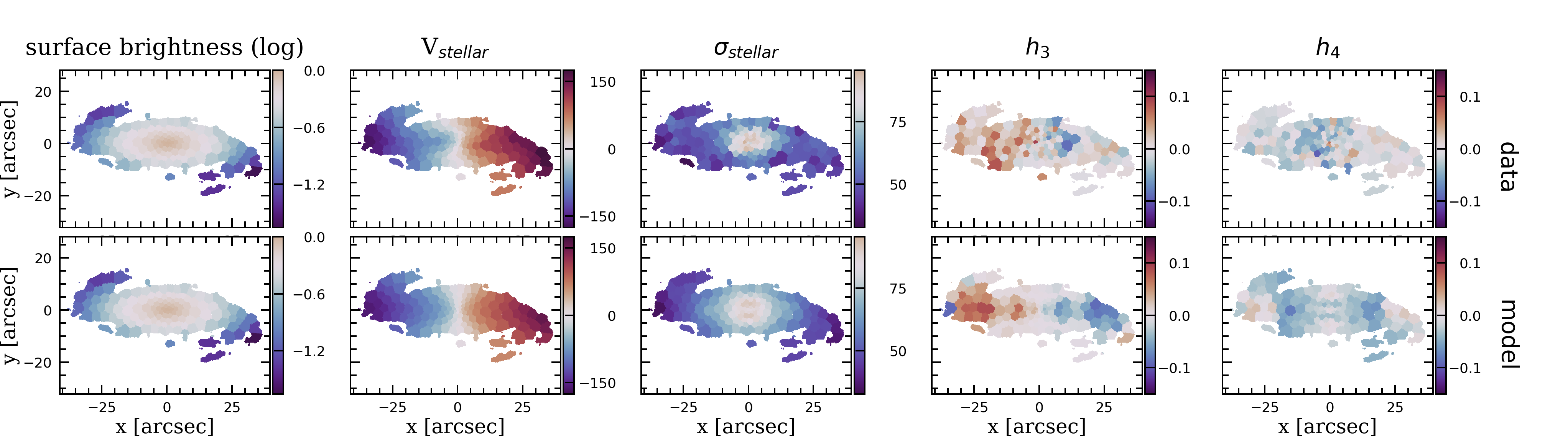}
    \caption{Same as \autoref{fig:kinmap_ESO339-G011}, but for NGC\,6754.}
    \label{fig:kinmap_NGC6754}
\end{figure*}

\begin{figure*}[p]
    \centering
    \includegraphics[width=\hsize]{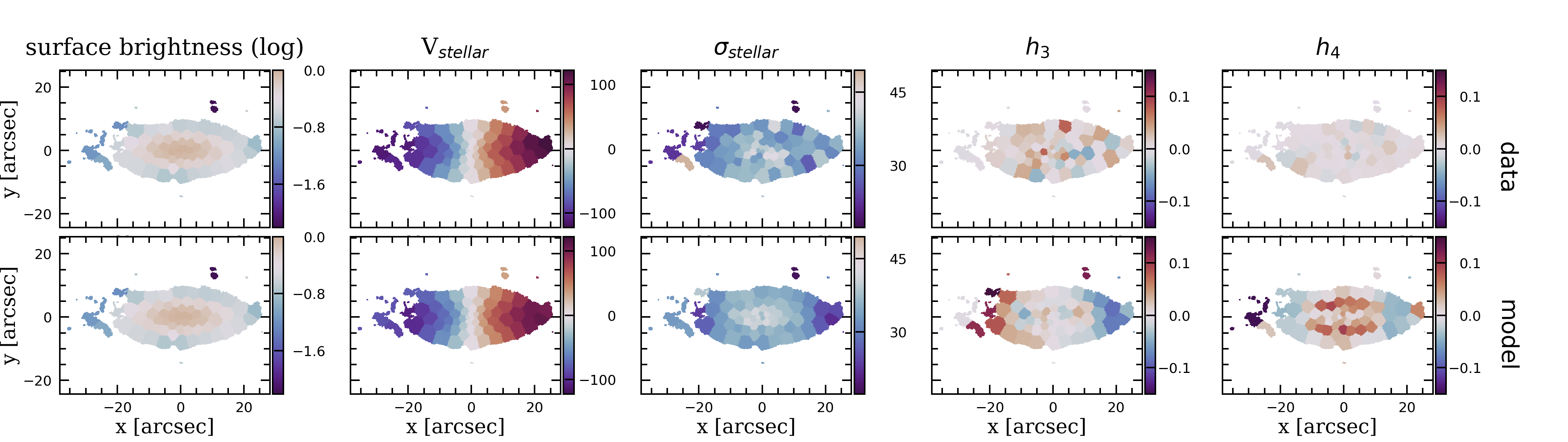}
    \caption{Same as \autoref{fig:kinmap_ESO339-G011}, but for NGC\,7162.}
    \label{fig:kinmap_NGC7162}
\end{figure*}

\begin{figure*}[p]
    \centering
    \includegraphics[width=\hsize]{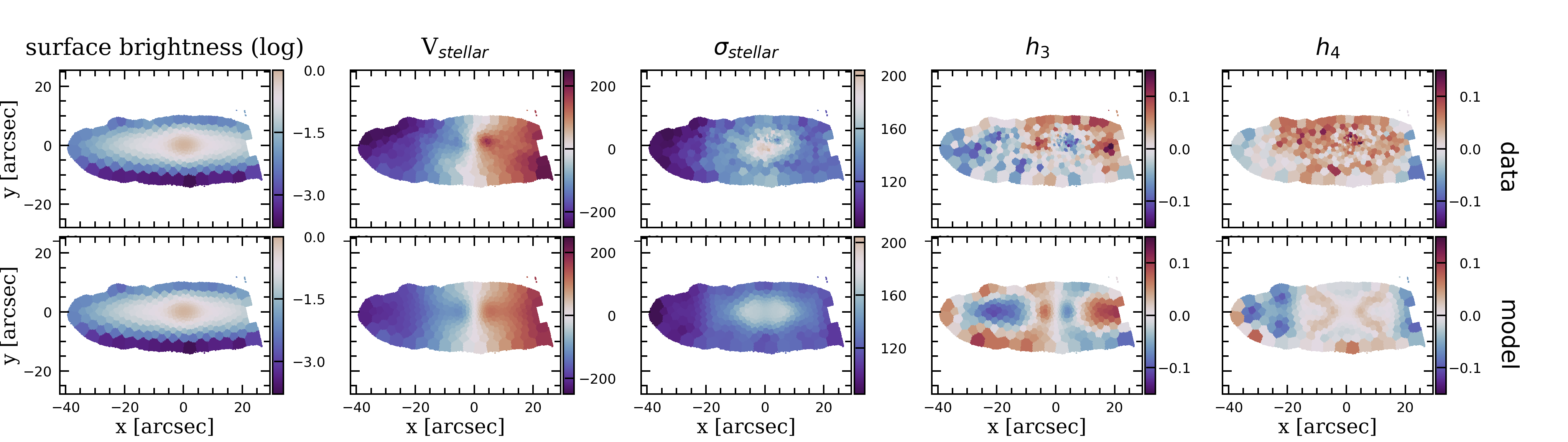}
    \caption{Same as \autoref{fig:kinmap_ESO339-G011}, but for NGC\,7771.}
    \label{fig:kinmap_NGC7771}
\end{figure*}

\end{appendix}

\end{document}